\documentclass[floatfix,aps,prd,reprint,preprintnumbers,groupedaddress,nofootinbib,nolongbibliography]{revtex4-2}
\usepackage{graphicx}

\input{epsf}

\usepackage{amssymb}
\usepackage{amsthm}

\usepackage{xcolor}

\usepackage{mathtools}
\usepackage{amsfonts}

\def\nn {\nonumber\\}
\def\ba {\begin{eqnarray}}
\def\ea {\end{eqnarray}}

\begin{document}
\preprint{JLAB-PHY-23-3785}

\title{Exact and Leading Order Radiative Effects in Semi-inclusive Deep Inelastic Scattering}

\author{Igor Akushevich}
\email[]{igor.akushevich@duke.edu}
\author{Stanislav Srednyak}
\email[]{stan.sredn@gmail.com}
\affiliation{Physics Department, Duke University, Durham, North Carolina 27708, USA}
\author{Alexander Ilyichev}
\email[]{ily@hep.by}
\affiliation{Belorussian State University, Minsk, 220030, Belarus}
\affiliation{Institute for Nuclear Problems Belorussian State University, Minsk, 220006, Belarus}

\begin{abstract}
Radiative effects in semi-inclusive hadron leptoproduction of unpolarized particles are calculated within the leading order approximation. The contributions of the infrared-free sum of the effects of real and virtual photon emission as well as the contribution of exclusive radiative tail are considered. It is demonstrated how the obtained formulae in the leading log approximation can be obtained using the standard approach of the leading log approximation as well as from the exact expressions for the radiative correction of the lowest order. The method of the electron structure function is used to calculate the higher order corrections. The results are analytically compared to the results obtained by other groups. Numeric illustrations are given in the kinematics of the modern experiments at Jefferson Laboratory.
\end{abstract}



\maketitle

\section{Introduction}

\begin{figure}[t]\centering
\scalebox{0.47}{\includegraphics{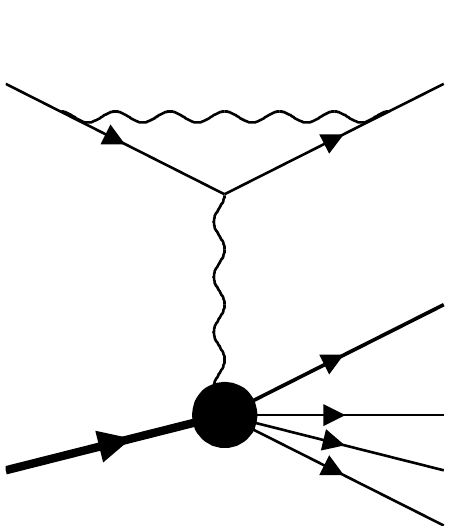}}
\scalebox{0.47}{\includegraphics{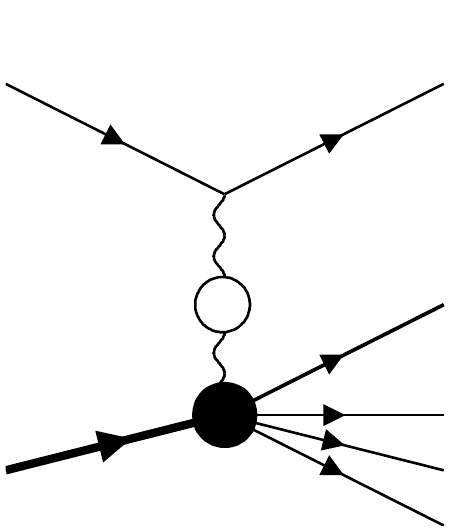}}
\\[-0.1cm]
{\bf \hspace{-.5cm} a) \hspace{3.3cm} b)\hspace{2.42cm}}
\\[0.1cm]
\scalebox{0.47}{\includegraphics{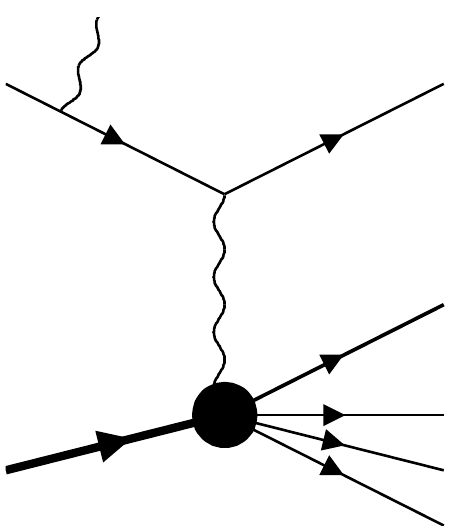}}
\scalebox{0.47}{\includegraphics{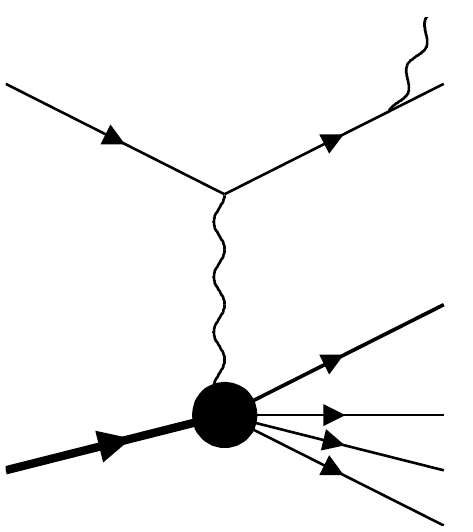}}
\\[-0.1cm]
{\bf \hspace{-.5cm} c) \hspace{3.3cm} d)\hspace{2.42cm}}
\\[0.1cm]
\scalebox{0.47}{\includegraphics{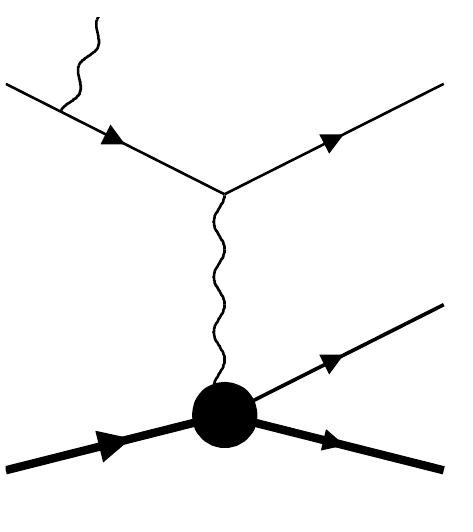}}
\scalebox{0.47}{\includegraphics{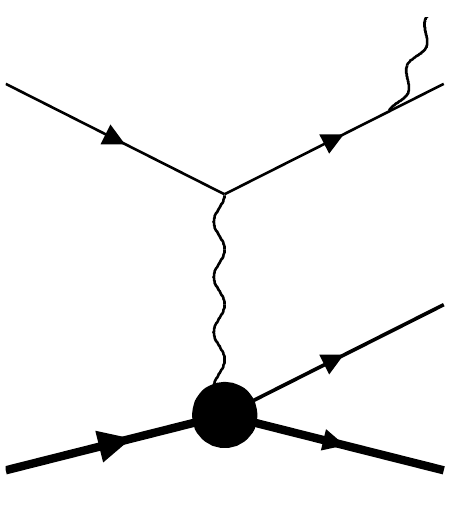}}
\\[-0.1cm]
{\bf \hspace{-.5cm} e) \hspace{3.3cm} f)\hspace{2.42cm}}
\\[0.1cm]
\caption{Feynman graphs for the contributions to the lowest order RC from semi-inclusive processes (a-d) and   exclusive  radiative tail
e) and f)}
\label{fgrc}
\end{figure}

Modern experiments on semi-inclusive deep inelastic in $ep$-scattering (SIDIS) provide information about multidimensional structure of the nucleon that is not  accessible in inclusive DIS.  Current and planned experiments in several laboratories, such as JLab, BNL,  and CERN have precision that necessitates consideration and implementation of radiative corrections (RC). The main contribution to  RC in SIDIS comes from emission of real photons by the initial and final electrons. The radiated photon is not detected in the detector by the design of SIDIS measurements, therefore the observed cross sections have to be integrated with respect to the phase space of the radiated photons. The integration in the soft photon region (i.e., when the photon energy is small) cannot be completed because of the infrared divergence that cancel in the sum with the contributions of loop diagrams (e.g., the vertex function in the lowest order). A special procedure of covariant extraction and cancellation of the infrared divergence developed by Bardin and Shumeiko \cite{BSh} is usually applied. An attractive property of the approach is the lack of simplifying assumptions that make the obtained formulae non-exact and dependent on artificial parameters, like $\Delta$, minimal photon energy in Mo and Tsai formalism \cite{Mo-Tsai}. An additional contribution to RC is the radiative tail from the exclusive peak (or exclusive radiative tail) that is characterized by the radiated photon and a single hadron in the unobserved hadronic state. This process contributes to RC to SIDIS when the invariant mass of the radiated photon and unobserved hadron equals the mass of the unobserved hadronic state in the base SIDIS process. The complete set of Feynman diagrams that are needed to be considered to calculate the lowest order RC is shown in Fig.~\ref{fgrc}.

The original formalism for RC in SIDIS in the simple quark-parton model was suggested in \cite{SSh1,SSh2}, that was later implemented in POLRAD 2.0 as a patch SIRAD \cite{Polrad}. The formulae allowed for calculating RC for the three-dimensional SIDIS cross section averaged over polar angle and transverse momentum of the final hadron. The formalism was then generalized in \cite{ASSh} to allow the calculation for the five-dimensional SIDIS cross section in scattering of unpolarized particles. The exclusive radiative tail was firstly calculated in \cite{AIO}.

The general calculation of RC for polarized particles was recently performed in \cite{AI2019}, and the code for numeric calculation of RC to the SIDIS cross section of electron scattering   arbitrary polarized particles was created. This calculation provides the so-called exact computation of RC. By “exactly” calculated RC we understand the estimation of the lowest order RC contribution with
any predetermined accuracy. The structure of the dependence on the electron mass in RC cross section is:
\begin{equation}
\sigma_{RC}=Al_m+B+O(m^2/Q^2)
,
\label{lonlo}
\end{equation}
where $l_m=\log({Q^2}/{m^2})$, and $A$ and $B$ do not depend on the electron mass $m$. If only $A$ is kept in the formulae for RC, this is the leading log approximation which evaluates the contributions of photons radiated collinerly to the initial or final electrons.
If both contributions are kept (i.e., contained $A$ and $B$), this is the calculation with the
next-to-leading accuracy, practically equivalent to exact calculation.


The leading log approximation for calculation of RC was firstly suggested in QCD by Dokshitzer \cite{Dsp}, Gribov-Lipatov \cite{GLsp}, and  Altarelli-Parisi \cite{APsp}. How the approximation can be applied for the lepton current was demonstrated by 
De Rújula, Petronzio, and Savoy-Navarro \cite{DeRujula}.  The QCD-based approach was adapted for real photon emission at the first order ${\cal O}( \alpha l_m)$ by Bl\"umlein \cite{Blumlein}, second order ${\cal O}( (\alpha l_m)^2)$ by Kripfganz, Mohring and Spiesberger \cite{Spies}, third order ${\cal O}( (\alpha l_m)^3)$ by Skrzypek \cite{Skrzypek}, second order subleading term ${\cal O}( \alpha^2 l_m)$ and fifth order ${\cal O}( (\alpha l_m)^5)$ by Bl\"umlein and  Kawamura in \cite{Blumlein2} and \cite{Blumlein3} respectively.

On the other side the leading log formulae can be also extracted from the exact formulae. Traditionally, such calculation represents a reasonable step in obtaining the formulae for RC  (e.g., exact \cite{dvcsrc1} and leading log \cite{dvcsrc2} formulae for RC to DVCS cross section) because the obtained formulae are compact and provide actually leading contribution of RC to the cross section. 

At last, complete resummation of the leading log terms in all orders in respect to $\alpha$ has been performed by Kuraev and Fadin in their seminal work \cite{kuraev1}.  In collaboration with Merenkov they showed how subleading terms  in all order of $\alpha$ have to be accounted in their resummation scheme \cite{kuraev2}. Such scheme was applied for polarized DIS \cite{AAM2004,ESFRAD} and for initial state QED radiation aspects in data analyses of future $e^+ e^-$ colliders  \cite{frixione2022initial}. 

Thus, three approaches to extract the leading log contributions for the SIDIS cross section (i.e., to
calculate $A$) include: 
i)
extract the poles that correspond to radiation collinear to initial and final electron, integrate
over angles, and find the factorized form traditional for leading log calculations;
ii) 
use our exact formulae, collect all terms that result in leading log after integration over
photon angles, combine them into the final expression, 
and iii)
use the method of the electron structure functions \cite{AAM2004}. All these approaches are applied and discussed in our paper. Recently, a new factorized approach to SIDIS was suggested which treats QED and QCD radiation equally \cite{Liu:2021jfp}. The approach is similar to the methods of electron structure functions, and the results obtained admit analytical comparison with our formulae.

We introduce the set of kinematical variables and calculate the Born cross section in Section \ref{SectBorn}. The calculations of RC using the three approaches are presented in Section \ref{SectRC}. Both SIDIS RC and the contribution of the exclusive radiative tail are studied in the leading and next-to-leading approximations. Numeric estimates in the kinematical conditions of modern experiments at JLab are presented in Section \ref{SectNum}. The leading, next-to-leading, as well as higher order correction obtained using the electron structure functions are numerically compared. Section \ref{SectDiscConc} contains discussion of the obtained results, computational tricks, role of the results in data analyses of SIDIS experiments, and comparison with the results obtained in \cite{Liu:2021jfp}.

\section{Born cross section}\label{SectBorn}

The SIDIS process
\ba
l(k_1)+n(p)\rightarrow l(k_2)+h(p_h)+x(p_x)
\label{sidisprocess}
\ea
($k_1^2=k_2^2=m^2$, $p^2=M^2$, $p_h^2=m_h^2$), are traditionally described by the
set of kinematical variables
\begin{eqnarray}
&\displaystyle
x=-\frac{q^2}{2qp},\;
y=\frac{qp}{k_1p},\;
z=\frac{p_hp}{pq},\;
t=(q-p_h)^2,
\;
\phi_h.
\label{setvar}
\end{eqnarray}
Here $q=k_1-k_2$,
$\phi_h$ is the angle between
$({\bf k_1},{\bf k_2})$ and $({\bf q},{\bf p_h})$ planes.

In most analyses the transverse momentum of the detected hadron $p_t$ or its square is used instead of $t$. Their relationship is presented below in Eq. (\ref{setvar3}). Formally, the $p_t$ is the orthogonal part of the 3-vector ${\bf p_h}$
with respect to
${\bf q}$ in the lab frame.

The set of additional quantities are used to describe the Born cross section. So, the invariants dependent on lepton momenta are identical to those used in DIS:
\begin{eqnarray}
&S=2pk_1,\;
Q^2=-q^2,\;
X=2pk_2,\;
S_x=S-X,\;
\nonumber\\&
\lambda_S=S^2-4m^2M^2,\;
\lambda_X=X^2-4m^2M^2,\;
\nonumber\\&
S_p=S+X,\;
\lambda_Y=S_x^2+4M^2Q^2,\;
\nonumber\\&
\lambda_1=Q^2(SX-M^2Q^2)-m^2\lambda_Y,\;
\nonumber\\&
W^2=(p+q)^2=S_x-Q^2+M^2,
\label{setvar2a}
\end{eqnarray}
whereas involvement of the detected hadron generates a set of new invariants:
\begin{eqnarray}
&V_{1,2}=2k_{1,2}p_h,\;
V_+=\frac 1 2(V_1+ V_2),
\nonumber\\&
V_-=\frac 1 2(V_1- V_2)=\frac 1 2(m_h^2-Q^2-t),
\nonumber\\&
S^\prime=2k_1(p+q-p_h)=S-Q^2-V_1,
\nonumber\\&
X^\prime=2k_2(p+q-p_h)=X+Q^2-V_2,
\nonumber\\&
p_x^2=(p+q-p_h)^2=M^2+t+(1-z)S_x.
\nonumber\\&
\lambda_S^\prime=S^{\prime 2}-4m^2p_x^2,\;
\lambda_X^\prime=X^{\prime 2}-4m^2p_x^2.
\label{setvar2b}
\end{eqnarray}

Noninvariant variables, such as the energy
$p_{h0}$, longitudinal $p_l$, and transverse $p_t$ ($k_t$) three-momenta of the detected hadron
(the incoming or scattering lepton)
with respect to the virtual photon direction,
in the target rest frame are expressed in terms of
the above invariants:
\begin{eqnarray}
&&\displaystyle
p_{h0}=\frac{zS_x}{2M},\;
\nonumber\\
&&\displaystyle
p_l
=\frac{zS^2_x-4M^2V_-}{2M\sqrt{\lambda_Y}}
=\frac{zS^2_x+2M^2(t+Q^2-m_h^2)}{2M\sqrt{\lambda_Y}},
\nonumber\\[2mm]
&&\displaystyle
p_t=\sqrt{p_{h0}^2-p_l^2-m_h^2},
\nonumber\\
&&\displaystyle
k_t=\sqrt{\frac{\lambda_1}{\lambda_Y}}.
\label{setvar3}
\end{eqnarray}

As a result the quantities $V_{1,2}$ can be written through $\cos \phi_h$
and other variables defined in Eqs.~(\ref{setvar})-(\ref{setvar3}) as
\begin{eqnarray}
V_1&=&p_{h0}\frac{S}{M}-\frac{p_l(S S_x+2M^2Q^2)}{M\sqrt{\lambda_Y}}-2 p_tk_t\cos \phi_h,
\nonumber\\
V_2&=&p_{h0}\frac{X}{M}-\frac{p_l(X S_x-2M^2Q^2)}{M\sqrt{\lambda_Y}}-2 p_tk_t\cos \phi_h.
\nonumber\\
\label{v1v2}
\end{eqnarray}

From the other side
\begin{eqnarray}
\cos \phi_h=\frac{S_pS_x(z Q^2+V_-)-\lambda_YV_+}{2p_t\sqrt{\lambda_Y\lambda_1}}.
\label{cphih}
\end{eqnarray}

\begin{figure}[t]\centering
\scalebox{0.47}{\includegraphics{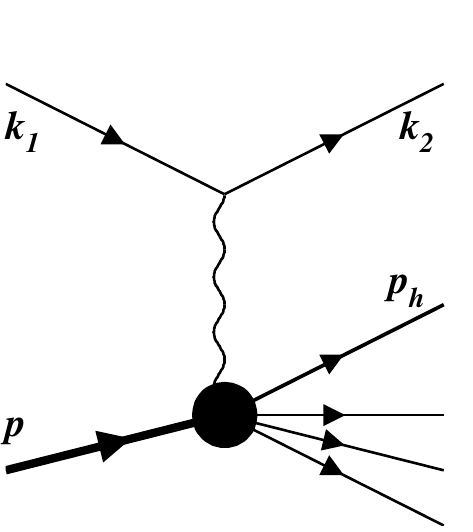}}
\caption{ Feynman graph for the lowest order SIDIS}
\label{fgb}
\end{figure}

The lowest order QED (Born) contribution to SIDIS is presented by the Feynman graph in Fig.~\ref{fgb}.
The cross section for this process reads
\ba
d\sigma _B=\frac{(4\pi\alpha)^2}{2SQ^4}W_{\mu \nu}(q,p,p_h)L^{\mu \nu}_B
d\Gamma _B,
\label{wl}
\ea
where the phase space is parameterized as
\ba
d\Gamma _B&=&
(2\pi)^4\frac{d^3k_2}{(2\pi)^32k_{20}}
\frac{d^3p_h}{(2\pi)^32p_{h0}}
\nn
&=&\frac1{4(2\pi)}\frac {S_xdx dy}{2} \frac{S_xdzdp^2_td\phi_h}{4Mp_l}.
\label{dgin}
\ea

The leptonic tensor can be presented as
\ba
L_B^{\mu \nu}&=&\frac 12 {\rm Tr}[({\hat k}_2+m)\gamma_{\mu }({\hat k}_1+m)\gamma_{\nu}]
\nn
&=&2k_{1}^{\mu}k_{2}^{\nu}+2k_{2}^{\mu}k_{1}^{\nu}-Q^2g^{\mu \nu}.
\label{ltborn}
\ea
According to \cite{AI2019} the hadronic tensor can be written in the covariant form
\ba
&\displaystyle
W_{\mu\nu}(q,p,p_h)=\sum\limits_{i=1}^4 w^i_{\mu\nu}(q,p,p_h){\cal H}_i=
-g^\bot_{\mu \nu} {\cal H}_1
\nn
&\displaystyle
+p^\bot_\mu p^\bot_\nu {\cal H}_2
+p^\bot_{h\mu} p^\bot_{h\nu} {\cal H}_3
+(p^\bot_{\mu} p^\bot_{h\nu}+p^\bot_{h\mu} p^\bot_{\nu}) {\cal H}_4
\label{ht1}
\ea
Here $g^\bot_{\mu \nu}=g_{\mu \nu}-q_{\mu }q_{\nu }/q^2$, for any four-vector $a^\bot_\mu=a_\mu+aq \; q_\mu/Q^2$.

Finally, we  find the Born contribution in the form
\ba
&\displaystyle
\sigma_B(S,Q^2,x,z,p_t,\cos \phi_h)\equiv \frac{d\sigma_B}{dxdydzdp_t^2d\phi_h}
\nn
&\displaystyle
=
\frac{\pi \alpha^2 S^2_x}{4MQ^4p_lS}
\sum\limits_{i=1}^4\theta^B_i(S,x,y,z,p_t,\cos \phi_h)
\nn
&\displaystyle
\times
{\cal H}_i(Q^2,x,z,p_t),
\label{born2}
\ea
where $\theta^B_i=L^{\mu \nu}_Bw^i_{\mu \nu}/2$,
\ba
\theta^B_1&=&Q^2,
\nonumber\\[1mm]
\theta^B_2&=&(SX-M^2Q^2)/2,
\nonumber\\[1mm]
\theta^B_3&=&(V_1V_2-m_h^2Q^2)/2,
\nonumber\\[1mm]
\theta^B_4&=&(SV_2+XV_1-zQ^2S_x)/2.
\label{thb}
\ea

The generalized structure functions can be expressed in terms of another set of the structure functions \cite{Bacchetta:2006tn} $F_{UU,T}$, $F_{UU,L}$,  
$F_{UU}^{\cos \phi_h}$ and $F_{UU}^{\cos 2\phi_h}$: 
\ba
{\cal H}_1&=&C_1[F_{UU,T}-F_{UU}^{\cos 2\phi_h}],\;
\nn
{\cal H}_2&=&\frac{4C_1}{\lambda_Y^2p_t^2}\biggl[\lambda_Yp_t^2Q^2F_{UU,L}
+\lambda_3^2S_x^2(F_{UU}^{\cos 2\phi_h}
+F_{UU,T})
\nn&&
-\lambda_2\lambda_Y(F_{UU,T}-F_{UU}^{\cos 2\phi_h}
)
\nn&&
+2S_x\lambda_3p_tQ\sqrt\lambda_YF_{UU}^{\cos \phi_h}
\biggr],
\nn
{\cal H}_3&=&\frac{2C_1}{p_t^2}
F_{UU}^{\cos 2\phi_h},
\nn
{\cal H}_4&=&-\frac{2C_1}{\lambda_Y p_t^2}[
2\lambda_3S_x
F_{UU}^{\cos 2\phi_h}
+p_tQ\sqrt{\lambda_Y}F_{UU}^{\cos \phi_h}
],
\ea
where $C_1=4Mp_l(Q^2+2xM^2)/Q^4$, $\lambda_2=V^2_-+m_h^2Q^2$, $\lambda_3=V_-+zQ^2$.
 The Born cross section (\ref{born2}) expressed in the terms of these structure functions has a rather simple structure, 
\ba
&\displaystyle
\sigma_B=\frac{\pi \alpha^2}{xQ^2}\frac y{1-\varepsilon}\biggl(1+\frac{\gamma^2}{2x}\biggl)\biggl\{
F_{UU,T}
+\varepsilon F_{UU,L}
\nn 
&\displaystyle
+\sqrt{2\varepsilon(1+\varepsilon)}\cos\phi_h F_{UU}^{\cos \phi_h}
+\varepsilon\cos 2\phi_h F_{UU}^{\cos 2\phi_h}
\biggr\},
\ea 
where $\gamma=2Mx/Q$ and $\varepsilon$ is the ratio of the longitudinal and transverse photon fluxes,
\ba
\varepsilon=\frac{1-y-\gamma ^2y^2/4}{1-y+y^2/2+\gamma ^2y^2/4}.
\ea

\section{Three approaches for leading logarithmic extraction}
\label{SectRC}

The QED RC come from three principal contributions: loop diagrams [Figs.~\ref{fgrc}(a) and~\ref{fgrc}(b)] and emission of the unobserved real photon in semi-inclusive [Figs.~\ref{fgrc}(c) and~\ref{fgrc}(d)] and exclusive [Figs.~\ref{fgrc}(e) and~\ref{fgrc}(f)] processes. The calculation of the loop diagrams involves the procedure of subtraction of the ultraviolet divergence which is based on the idea of the electric charge renormalization. After that the integral over loop momentum still contains the infrared divergence that cancels in the sum with the contribution of the real photon emission in a semi-inclusive process. The contribution of the exclusive radiative tail does not contain the infrared divergence because of kinematical restrictions and can be calculated separately from other contributions.

As it was mention in Introduction there are three approaches to extracting leading log contributions. In this section we describe each of them 
\subsection{Extraction of the Collinear Poles}
\label{subsecCollpoles}
The contribution of real photon emission
\ba
l(k_1)+n(p)\rightarrow l(k_2)+h(p_h)+x(p_x)+\gamma(k)
\label{sidisrprocess}
\ea
($k^2=0$) from the lepton leg shown in Fig.~\ref{fgrc}(a), \ref{fgrc}(b), can be presented as a convolution of
the leptonic tensor with the real photon emission whose structure is well-known:
\ba
L_{R}^{\mu \nu }&=&-\frac 12 {\rm Tr}[({\hat k}_2+m)\Gamma^{\mu \alpha}_R
({\hat k}_1+m){\bar \Gamma}^{\nu}_{R\alpha}],
\nn
\Gamma^{\mu \alpha }_R&=&\biggl(\frac{k_1^{\alpha}}{kk_1}-\frac{k_2^{\alpha}}{kk_2} \biggr)\gamma ^\mu
-\frac{\gamma ^\mu{\hat k}\gamma ^{\alpha }}{2kk_1}-\frac{\gamma ^\alpha {\hat k}\gamma ^{\mu }}{2kk_2},
\nonumber\\
{\bar \Gamma^{\nu}_{R\alpha }}&=&\gamma_0\Gamma^{\nu \dagger}_{R \alpha }\gamma_0
\nonumber\\&=&
\biggl(\frac{k_{1\alpha}}{kk_1}-\frac{k_{2\alpha}}{kk_2} \biggr)\gamma ^\nu
-\frac{\gamma ^\nu{\hat k}\gamma _{\alpha }}{2kk_2}-\frac{\gamma _\alpha {\hat k}\gamma ^{\nu }}{2kk_1},
\label{lr}
\ea
and hadronic tensor (\ref{ht1}):
\ba
d\sigma^R=\frac{(4\pi\alpha)^3}{2S(q-k)^4}W_{\mu \nu}(q-k,p,p_h)L_{R}^{\mu \nu}d\Gamma_R,
\label{dsr}
\ea
where  
\ba
d\Gamma_R&=&(2\pi)^4\frac{d^3k}{(2\pi)^32k_{0}}
\frac{d^3k_2}{(2\pi)^32k_{20}}
\frac{d^3p_h}{(2\pi)^32p_{h0}}
\nn [2mm]
&=&\frac1{8(2\pi)^4}\frac {S S_xdx dy }{2\sqrt{\lambda_S}} \frac{S_xdzdp^2_td\phi_h}{4Mp_l}\frac{d^3k}{k_0}.
\label{dgr}
\ea

Integration over the photon angles can result in the leading log term. For example, 
\ba
\int  \frac{d\Omega_k}{kk_1}&=&\frac{1}{E_\gamma}\int  \frac{d\Omega_k}{E_1-|\bf{k}_1|\cos\theta_\gamma} 
\nn[2mm]
&=&\frac{2\pi}{E_\gamma |\bf{k}_1|}\log\frac{E_1+|\bf{k}_1|}{E_1-|\bf{k}_1|}
\approx \frac{2\pi}{E_\gamma E_1}\log\frac{4E_1^2}{m^2}.
\nn  
\ea
Similarly, integration of the terms with $(kk_1)^{-2}$ results in:
\ba
\int  \frac{d\Omega_k}{(kk_1)^2}=\frac{1}{E_\gamma^2}\int  \frac{d\Omega_k}{(E_1-|\bf{k}_1|\cos\theta_\gamma)^2} 
\approx\frac{2\pi}{E_\gamma^2 E_1^2 }\frac{2E_1}{m^2}.
\nn
\ea
Since the squared propagators appear with a factor of $m^2$, i.e., as $m^2/(kk_1)^2$ and $m^2/(kk_2)^2$, such terms do not result in the leading log terms. Thus, the procedure of extraction of leading log term in the standard leading log approximation \cite{DeRujula, Blumlein,Spies,AI2012,dvcsrc2} contains the following steps. 
In each convolution of leptonic tensor $L_R^{\mu \nu}$  with the tensor structures $w^i_{\mu \nu}$ in the hadronic tensor, the electron mass can be neglected everywhere in the numerators.
Then, the terms containing $1/kk_1$ and $1/kk_2$ have to be extracted, i.e., the convolutions have to be presented in the form of two terms, that are historically known as $s$- and $p$-peaks
\ba
&\displaystyle
L_R^{\mu\nu}w^i_{\mu\nu}(q-k,p,p_h){\cal H}_i(q-k)=\frac{G^i_s(k,...){\cal H}_i(q-k)}{kk_1}
\nn
&\displaystyle
+ \frac{G^i_p(k,...){\cal H}_i(q-k)}{kk_2}.
\label{GsGp}
\ea
We note, that the convolutions can have the terms with $1/(kk_1\;\;kk_2$) that can be decomposed as  
\ba
\frac{1}{kk_1}\frac{1}{kk_2}=-\frac{1}{k(k_1-k_2)}\frac{1}{kk_1}+\frac{1}{k(k_1-k_2)}\frac{1}{kk_2}.
\ea
The term $k(k_1-k_2)$ is regular (i.e., not equaling zero for any peak) and can be included to a respective $G^i_{s,p}(k,...)$. 
Since $G^i_{s,p}$ are regular functions of the momentum $k$, this momentum (as well as all kinematical variables containing $k$) can be taken in the respective peaks in $G^i_{s,p}$ as well as in arguments of structure functions ${\cal H}_i$. The four arguments of structure functions ${\cal H}_i$ come from the four scalar products $pq$, $pp_h$, $q^2$, and $qp_h$. Only the vector $q$ has to change if the photon is radiated $q\rightarrow q-k$.  Therefore, we can write for (\ref{GsGp}):
\ba
&\displaystyle
\frac{L_R^{\mu\nu}w^i_{\mu\nu}(q-k,p,p_h){\cal H}_i(q-k)}{(q-k)^2}
\nn 
&\displaystyle
=
\frac{G_s^i(k_s,...){\cal H}_i(q-k_s)}{(q-k_s)^2kk_1}
+ \frac{G_p^i(k_p,...){\cal H}_i(q-k_p)}{(q-k_p)^2kk_2}.
\label{GsGp1}
\ea
 In the standard leading log approximation the substitutions of the vector $k$ in the $s$- and $p$-peaks are performed by introduction of dimensionless variables $z_1$ and $z_2$, that reflect remaining degree of freedom, i.e., photon energy,  as follows $k\to k_{s,p}$ where
\ba
k_s=(1-z_1)k_1,\;
k_p=(z_2^{-1}-1)k_2
\label{kpar}
\ea
for $s$-, $p$-peaks respectively. 
The possibility to substitute $k$ in $G^i_{s,p}$ is justified by the fact that the difference $G^i_s(k,...)-G^i_s(k_s,...)$ is exactly zero in the peak respective integration of this difference divided by $kk_1$ does not result in the leading log. 

The integration of (\ref{GsGp1}) over angular variables can be formally presented as: 
\ba
\frac {d^3k}{k_0}\frac 1{k_1k}=2\pi l   _m dz_1,\qquad
\frac {d^3k}{k_0}\frac 1{k_2k}=2\pi l_m \frac{dz_2}{z_2^2},
\label{k12}
\ea
where
\ba
l_m=\log \frac{Q^2}{m^2}.
\ea

The above procedure can be formalized in terms of leptonic tensor (\ref{lr}), which split in two respective parts in the leading log approximation: 
\ba
L_{Rs}^{\mu \nu }&=&\frac{1+z_1^2}{z_1(1-z_1)}\frac 1{k_1k}L_{B}^{\mu \nu }(k_1\to z_1k_1),
\nonumber\\
L_{Rp}^{\mu \nu }&=&\frac{1+z_2^2}{1-z_2}\frac 1{k_2k}L_{B}^{\mu \nu }(k_2\to k_2/z_2).
\label{lrsp}
\ea
The convolution with the hadronic tensor is:
\begin{gather}
L^{\mu \nu }_RW_{\mu\nu}(q-k,p,p_h)=L^{\mu \nu }_{Rs}W_{\mu\nu}(q-k_s,p,p_h)+\\ \nonumber
+L^{\mu \nu }_{Rp}W_{\mu\nu}(q-k_p,p,p_h).
\end{gather}

\begin{figure}[t!]
\includegraphics[width=8.5cm,height=7cm]{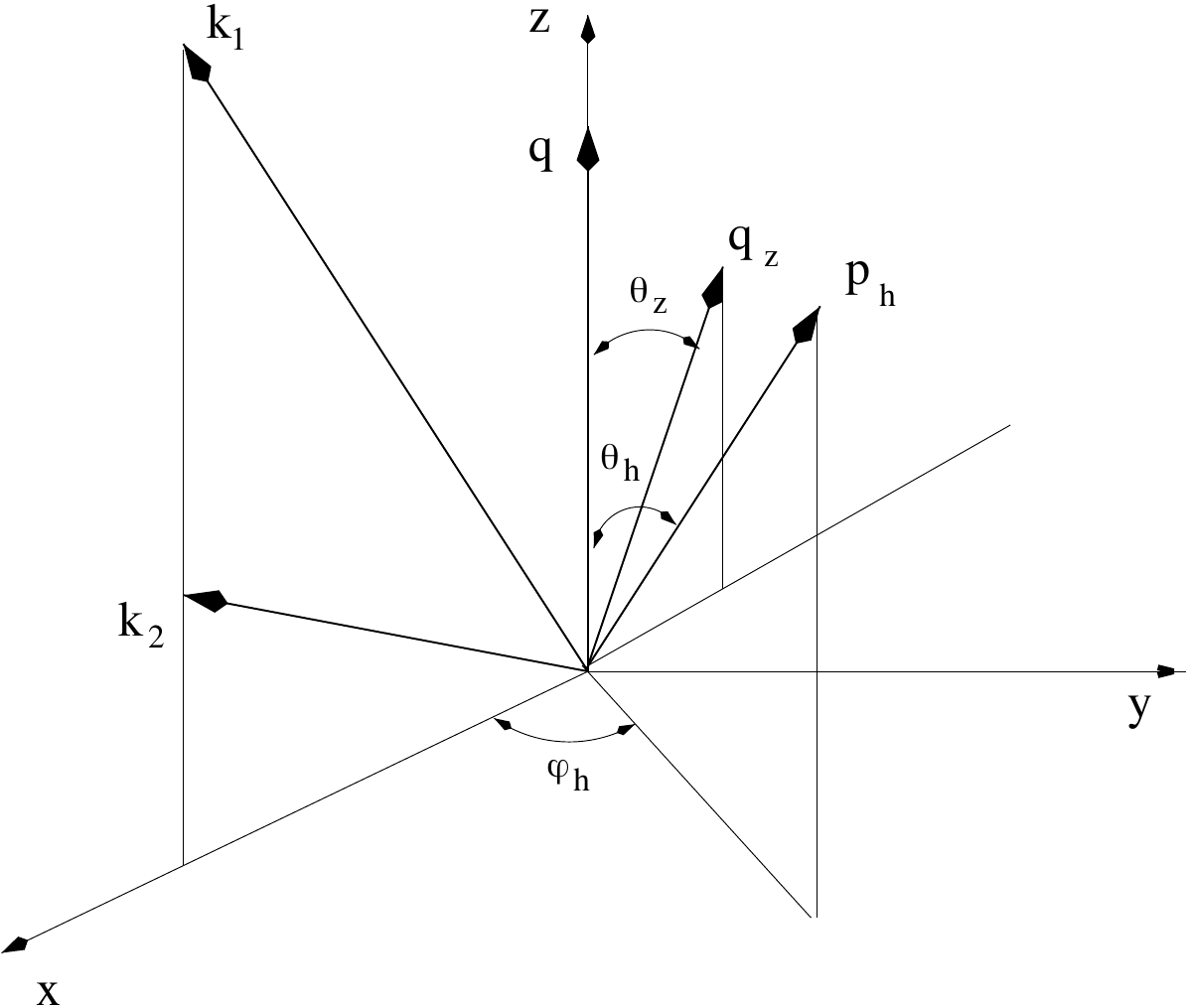}
       \caption{The momenta of the particles of SIDIS process (\ref{sidisprocess}) in the lab. frame; $q$ and $q_z$ are the momenta of the virtual photon in the original and shifted kinematics.}
\label{vect}
\end{figure}

This approach has a useful geometric interpretation. We see that the matrix element squared is calculated as convolution of Born leptonic tensor with a shifted
momentum of initial (or final) electron for $s$-(and $p$-) peaks. This means the parametrization (\ref{kpar}) allows to
write collinear bremsstrahlung in terms of the Born cross section but in a so-called shifted born condition
\ba
z_1k_1+p=k_2+p_h+p_x,\;
\nn
k_1+p=k_2/z_2+p_h+p_x.\;
\label{kpar1}
\ea
The kinematics of the process is sketched in Fig.~\ref{vect}. The momentum transfer $q=k_1-k_2$ is chosen along the axis $z$, and vectors $k_1,k_2$ constitute $x,z$-plane. This  fixes the coordinate system. In the leading approximation $k \rightarrow (1-z_1)k_1$ or $k \rightarrow (1/z_2-1)k_2$ lies entirely in the $xz$-plane.

After substitution of (\ref{dgr},\ref{lrsp},\ref{kpar}) into (\ref{dsr}) and
taking into account the angular integration of the first-order poles (\ref{k12})
one can find that the leading order approximation of the real photon emission to $s$- and $p$-peaks can be expressed through the Born
contribution  $\sigma_B$ (\ref{born2}) with so-called shifted variables in a following way:
\ba
d\sigma^{in\;s}_{1L}&=&\frac\alpha {2\pi }l_m dz_1\frac{1+z_1^2}{1-z_1}\frac{p_{ls}S_x^2}{p_l(z_1S-X)^2}
\nonumber\\&&\times
\sigma_B(z_1S,z_1Q^2,x_s,z_s,p_{ts},\cos \phi_{hs}),
\nonumber\\
d\sigma^{in\;p}_{1L}&=&\frac\alpha {2\pi }l_m dz_2\frac{1+z_2^2}{z_2^2(1-z_2)}\frac{p_{lp}S_x^2}{p_l(S-X/z_2)^2}
\nonumber\\&&\times
\sigma_B(S,z_2^{-1}Q^2,x_p,z_p,p_{tp},\cos \phi_{hp}).
\label{dsrsp}
\ea

The quantities with the subscripts $s$ and $p$ read:
\ba
&\displaystyle
x_s=\frac{z_1Q^2}{(z_1S-X)},\;
z_s=\frac{zS_x}{(z_1S-X)},\;
\nonumber
\ea
\vspace{-6mm}
\ba
&\displaystyle
\lambda_{Ys}=(z_1S-X)^2+4z_1M^2Q^2,
\nonumber
\ea
\vspace{-6mm}
\ba
&\displaystyle
p_{ls}=\frac{z S_x(z_1S-X)-2M^2(z_1V_1-V_2)}{2M\sqrt{\lambda_{Ys}}},\;
\nonumber
\ea
\vspace{-6mm}
\ba
&\displaystyle
p_{ts}=\sqrt{\frac{z^2S_x^2}{4M^2}-p_{ls}^2-m_h^2},
\nonumber
\ea
\vspace{-6mm}
\ba
&\displaystyle
\cos \phi_{hs}=\frac{1}{4z_1p_{ts}\sqrt{\lambda_{Ys}\lambda_1}}\Biggl[(z_1S+X)(2z_1zS_x Q^2
\nonumber
\ea
\vspace{-6mm}
\ba
&\displaystyle
+(z_1S-X)(z_1V_1-V_2))-\lambda_{Ys}(z_1V_1+V_2)\Biggr],
\nonumber
\ea
\vspace{-6mm}
\ba
&\displaystyle
x_p=\frac{Q^2}{(z_2S-X)},\;
z_p=\frac{zS_x}{(S-z_2^{-1}X)},\;
\nonumber
\ea
\vspace{-6mm}
\ba
&\displaystyle
\lambda_{Yp}=(S-z_2^{-1}X)^2+4z_2^{-1}M^2Q^2,
\nonumber
\ea
\vspace{-6mm}
\ba
&\displaystyle
p_{lp}=\frac{z S_x(S-z_2^{-1}X)-2M^2(V_1-z_2^{-1}V_2)}{2M\sqrt{\lambda_{Yp}}},\;
\nonumber
\ea
\vspace{-6mm}
\ba
&\displaystyle
p_{tp}=\sqrt{\frac{z^2S_x^2}{4M^2}-p_{lp}^2-m_h^2},
\nonumber
\ea
\vspace{-6mm}
\ba
&\displaystyle
\cos \phi_{hp}=\frac{z_2}{4p_{tp}\sqrt{\lambda_{Yp}\lambda_1}}\Biggl[(S+z_2^{-1}X)(2z_2^{-1}zS_x Q^2
\nonumber
\ea
\vspace{-6mm}
\ba
&\displaystyle
+(S-z_2^{-1}X)(V_1-z_2^{-1}V_2))-\lambda_{Yp}(V_1+z_2^{-1}V_2)\Biggr].
\nn
\label{shift}
\ea

The expressions (\ref{dsrsp}) are infrared divergent  at $z_{1,2}\to~1$. This infrared divergence is canceled in the sum with the contribution from the vertex function  presented by Feynman graph in Fig.~\ref{fgrc}(a).

In the leading log approximation the incorporation of the vertex function contribution and respective cancellation of the infrared divergence is implemented using the electron splitting function, which was originally suggested for the use in QCD \cite{Dsp,GLsp,APsp} and then was adapted for real photon emission \cite{Blumlein,Spies}.
The splitting function is defined through the so-called (+)-operator, 
\ba
P(z)=\frac{1+z^2}{(1-z)_+}, 
\label{spf}
\\\nonumber
\ea
and use to replace the factor $(1+z^2)/(1-z)$ in the leading log formulae. This 
(+)-operator is defined as
\ba
\int\limits_x^1dzP(z)f(z)=\int\limits_x^1dz\frac{1+z^2}{1-z}(f(z)-f(1))
\nn
-f(1)\int\limits_0^xdz\frac{1+z^2}{1-z}.
\label{spf1}
\ea  

Application of the splitting function to equations (\ref{dsrsp}) leads 
to
\ba
\sigma^{in\;s}_{1L}&=&\frac\alpha {2\pi }l_m \int\limits_{z_{1i}}^1dz_1P(z_1)\frac{p_{ls}S_x^2}{p_l(z_1S-X)^2}
\nonumber\\&&\times
\sigma_B(z_1S,z_1Q^2,x_s,z_s,p_{ts},\cos \phi_{hs}),
\nonumber\\
\sigma^{in\;p}_{1L}&=&\frac\alpha {2\pi }l_m \int\limits_{z_{2i}}^1\frac{dz_2}{z_2^2}P(z_2)\frac{p_{lp}S_x^2}{p_l(S-X/z_2)^2}
\nonumber\\&&\times
\sigma_B(S,z_2^{-1}Q^2,x_p,z_p,p_{tp},\cos \phi_{hp}).
\\\nonumber
\label{dsrsp1}
\ea
The lowest limits of integration can be found through SIDIS pion threshold
\ba
z_{1i}&=&1-(p_x^2-M_{th}^2)/S^\prime,\nn
z_{2i}&=&\frac 1{(1+(p_x^2-M_{th}^2)/X^\prime)}
.
\label{z12m}
\ea
Here $M_{th}$ is the minimal value of the invariant mass of the undetected hadrons $p_x$ for the SIDIS process, e.g., $M_{th}=M+m_\pi$ when the detected hadron is the pion.

The final expression for the RC in SIDIS in the leading log approximation is:
\begin{widetext}
\ba
\sigma^{in}_{1L}
&=&\Biggl[1+
\frac{\alpha }{\pi }\delta_{\rm vac}^l(Q^2)\biggr]\sigma_B(S,Q^2,x,z,p_t,\phi_h)+\sigma^{in\;s}_{1L}+\sigma^{in\;p}_{1L}
\nonumber \\
&=&\Biggl[1+
\frac{\alpha }{\pi }\delta_{\rm vac}^l(Q^2)\biggr]\sigma_B(S,Q^2,x,z,p_t,\phi_h)
\nonumber \\&&
+\frac{\alpha}{2\pi}l_m\int\limits_0^1 dz_1
\frac{1+z_1^2}{1-z_1}\Biggl[\theta(z_1-z_{1i})
\frac{p_{ls}S_x^2}{p_lS_{xs}^2}
\sigma_B(z_1S,z_1Q^2,x_s,z_s,p_{ts},\cos \phi_{hs})
-
\sigma_B(S,Q^2,x,z,p_t,\cos \phi_h)
\Biggr]
\nonumber \\&&
+\frac{\alpha}{2\pi}l_m\int\limits_0^1 dz_2
\frac{1+z_2^2}{1-z_2}\Biggl[\frac{\theta(z_2-z_{2i})}{z_2^2}\frac{p_{lp}S_x^2}{p_lS_{xp}^2}
\sigma_B(S,z_2^{-1}Q^2,x_p,z_p,p_{tp},\cos \phi_{hp})
-
\sigma_B(S,Q^2,x,z,p_t,\cos \phi_h)
\Biggr]
\label{sll1}
\ea
\end{widetext}
Here we added the contribution of vacuum polarization by electron [Fig.~\ref{fgrc}(b)] in leading approximation which is external to the approach involving the splitting function and has to be added separately: 
\ba
\delta_{\rm vac}^l(Q^2)=\frac 2 3l_m.
\label{dvac}
\ea
A direct proof that the splitting function works for 
SIDIS is presented in Appendix~\ref{app}.

Similar calculation can be applied for extracting the leading approximation from the exclusive radiative tail depicted in Figs. \ref{fgrc}(e) and \ref{fgrc}(g)
\ba
l(k_1)+n(p)\rightarrow l(k_2)+h(p_h)+u(p_u)+\gamma(k),
\label{excprocess}
\ea
where $p_u$ is the four-momentum of a single undetected
hadron ($p_u^2 = m^2_u$). This process gives the contribution to SIDIS because the  mass square of the undetected particles $(p_u+k)^2$ can exceed the pion threshold $M_{th}^2$ for the rather hard photon emission. As a result the exclusive radiative tail does not contain infrared divergence. Moreover, since the fifth SIDIS variable $z$ is fixed by the energy of the emitted photon in leading approximation photonic variables $z_{1,2}$ are also fixed.  
The explicit expression for  
the exclusive radiative tail in the leading approximation presented below by Eq.~(\ref{sexll1}) of Subsection~\ref{ert}.

\subsection{Extraction leading log correction from  exact equations}
\label{llext}
The expression for the lowest order RC calculated exactly in \cite{AI2019} is:
\ba
\sigma^{in}&=&\Biggl[1+
\frac{\alpha }{\pi }(\delta_{VR}
+\delta_{\rm vac}^l
+\delta_{\rm vac}^h
)\biggr]\sigma^{B}(S,Q^2,x,z,p_t,\phi_h)          
\nn&&
+\sigma^F_R
+\sigma^{\rm AMM}.
\label{srv}
\ea
Two quantities, $\delta_{\rm vac}^h$ ($\sigma^{\rm AMM}$), do not contribute to RC in the leading approximation, since $\delta_{\rm vac}^h$ is independent on the electron mass and $\sigma^{\rm AMM}$ is proportional to it. The expressions for $\delta_{VR}$ and $\delta_{\rm vac}^l$ in the leading approximation are presented in Eqs.~(\ref{dvrll}) and (\ref{dvac}) respectively.

The exact expression for $\sigma^F_R$ is defined by Eq.~(43) of \cite{AI2019}
through the integration over three photonic variables
\ba
R=2kp,\;\tau=kq/kp,\; \phi_k,
\ea
where $\phi_k$ is
an angle between $({\bf k}_1,{\bf k}_2)$ and $({\bf k},{\bf q})$ planes
\ba
\sigma_R^F&=&-\frac{\alpha ^3S S_x^2}{32        \pi Mp_l\lambda_S\sqrt{\lambda _Y}
}
\int\limits_{\tau _{\rm min}}^{\tau _{\rm max}}
d\tau  
\int\limits_{0}^{2\pi}
d\phi_k  
\int\limits_{0}^{R _{\rm max}}
dR
\nn &&\times  
\sum\limits_{i=1}^4
\Biggl[
\sum\limits_{j=1}^3\frac{ {\cal H}_i(Q^2+\tau R,\tilde x,\tilde z, \tilde p_t)\theta^0_{ij}R^{j-2}}{(Q^2+\tau R)^2}
\nn &&
-
\frac{\theta^0_{i1}{\cal H}_i(Q^2,x,z,p_t)}{RQ^4}
\Biggr].
\label{srfin}
\ea
Here the variable with tilde are defined as
\ba
&\displaystyle
\tilde x=\frac{Q^2+\tau R}{S_x-R},\;
\tilde z=\frac{zS_x}{S_x-R},
\nn
&\displaystyle
\tilde p_t^2=\frac{z^2S_x^2}{4M^2}-\frac{(z S_x(S_x-R)-2M^2(2V_--\mu R))^2}{4M^2((S_x-R)^2+4M^2(Q^2+\tau R))}
\nn
&\displaystyle
-m_h^2.
\ea
and have meaning of the usual SIDIS variables in the shifted kinematics.
The limits of integrations are
\ba
R_{\rm max}=\frac {p_x^2-M_{th}^2}{1+\tau-\mu },
\;\;\;\;
\tau _{\rm max/min}=\frac{S_x\pm \sqrt{\lambda_Y}}{2M^2},
\ea
where
\ba
\mu&=&\frac{kp_h}{kp}=\frac{p_{h0}}{M}+\frac{p_l(2\tau M^2-S_x)}{M\sqrt{\lambda_Y}}
\nonumber\\&&
-2Mp_t\cos(\phi_h+\phi_k)\sqrt{\frac{(\tau_{\rm max}-\tau)(\tau-\tau_{\rm min})}{\lambda_Y}}.
\nn
\label{mudef}
\ea
After replacing variable $R$ by
\ba
R^\prime =(1+\tau-\mu) R,
\label{Rh}
\ea
the region of integration transform into cuboid
\ba
\int\limits_0^{R_{\rm max}}dR\to \int\limits_0^{p_x^2-M_{th}^2}\frac {dR^\prime}{1+\tau-\mu}, 
\label{subsRtoRp}
\ea
that allow us to perform the integration over $R^\prime$ as external.

The quantities $\theta_{ij}^0$ ($i=1,\dots,4$ and $j=1,\dots,3$) in (\ref{srfin}) are defined in Appendix B of  \cite{AI2019}. They result from convolution of the leptonic tensor (\ref{lr}) with hadronic structures $w^i_{\mu\nu}$. These quantities
contain the terms corresponding to $s$- and $p$-peaks, which are localized in (B.5) of \cite{AI2019} and can be presented in our notation as: 
\ba
F_d&=&\frac {R^2}{4 kk_1\; kk_2}=\frac 1\tau \Biggl(\frac R{2kk_2}-\frac R{2kk_1}\Biggr),
\nn
F_{1+}&=&\frac R{2kk_2}+\frac R{2kk_1},
\nn
F_{2\pm}&=&\frac {R^2}{4kk_2^2}\pm\frac {R^2}{4kk_1^2},
\ea
with
\ba
\frac {2k_ik}{R}&=&a_i+b\cos\phi_k,
\ea
where
\ba
a_1&=&\frac {Q^2S_p+\tau (SS_x+2M^2Q^2)}{\lambda_Y},
\nn
a_2&=&\frac {Q^2S_p+\tau (XS_x-2M^2Q^2)}{\lambda_Y},
\nonumber\\
b&=&-\frac {2M\sqrt{(\tau _{\rm max}-\tau)(\tau-\tau _{\rm min})\lambda_ 1}}{\lambda_Y}.
\label{z1z2}
\ea
The integration of these terms over $\phi_k$ give ($n=1,2$) 
\ba
\int\limits_0^{2\pi}\frac{d\phi_k}{a_n+b\cos\phi_k}&=&
\frac{2\pi }{\sqrt{C_n}},\nn
\int\limits_0^{2\pi}\frac{d\phi_k}{(a_n+b\cos\phi_k)^2}&=&
\frac{2\pi a_n}{C_n^{3/2}},
\label{z1z2i1}
\ea 
with
\ba
C_1&=&\frac{S^2(\tau-\tau_s)^2+4m^2M^2(\tau-\tau_{min})(\tau_{max}-\tau)}{\lambda _Y},
\nn
C_2&=&\frac{X^2(\tau-\tau_p)^2+4m^2M^2(\tau-\tau_{min})(\tau_{max}-\tau)}{\lambda _Y}.
\nn
\ea
Due to the smallness of the lepton mass the the expressions for $C_{1,2}$ have a sharp peak for $\tau \to \tau _s\equiv -Q^2/S $ and $\tau \to \tau _p\equiv Q^2/X $, respectively. Note the quantities $\tau_{s,p}$ can be also obtained from $\tau=kq/kp$ by the replacement $k\to k_{s,p}$ from (\ref{kpar}). 

The integration over $\tau$ of expressions (\ref{z1z2i1}) can be performed analytically, 
\ba
\int \limits_{\tau _{\rm min}}^{\tau _{\rm max}}d\tau \int \limits _0^{2\pi } \frac{d\phi_k}{a_1+b\cos\phi_k}
&=&2\pi \sqrt{\lambda _Y }L_S,
\nn[1mm]
\int \limits_{\tau _{\rm min}}^{\tau _{\rm max}}d\tau \int \limits _0^{2\pi } \frac{d\phi_k}{a_2+b\cos\phi_k}
&=&2\pi \sqrt{\lambda _Y }L_X.
\nn[1mm]
\int \limits_{\tau _{\rm min}}^{\tau _{\rm max}}d\tau \int \limits _0^{2\pi } \frac{d\phi_k}{(a_i+b\cos\phi_k)^2}
&=&\frac{2\pi
\sqrt{\lambda _Y }}{m^2},
\label{z1z2i2}
\ea
with
\ba
L_S&=&\frac 1{\sqrt{\lambda _S }}\log\frac{S+\sqrt{\lambda _S }}{S-\sqrt{\lambda _S }}
\nn
&=&\frac 1 S\biggl[ l_m+\log\frac{S^2}{Q^2M^2}\biggr]+{\cal O}\left(\frac{m^2}{Q^2}\right),
\nn
L_X&=&\frac 1{\sqrt{\lambda _X }}\log\frac{X+\sqrt{\lambda _X }}{X-\sqrt{\lambda _X }}
\nn
&=&\frac 1 X\biggl[ l_m+\log\frac{X^2}{Q^2M^2}\biggr]+{\cal O}\left(\frac{m^2}{Q^2}\right).
\label{lsx}
\ea
We see, that only the first order poles ($1/k_1k$ and $1/k_2k$) contribute to RC in the leading approximation.

Actually, the integrand in (\ref{z1z2i2}) depends on $\tau$ and $\phi_k$ not only in $\theta_{ij}^0$ but also in arguments of 
 structure functions ${\cal H}_i$, the photonic propagator squared $(Q^2+\tau R)^{-2}$, and the factor $(1+\tau-\mu)$ that appeared after the substitution of the integration variable  $R\to R^\prime $  in Eq.~(\ref{subsRtoRp}). All these functions are regular (i.e., equivalent neither zero nor infinity in the integration region). Therefore, we can make the identical transformation for extraction of the leading and next-to-leading terms:
\begin{eqnarray}
&\displaystyle
\int \limits_{\tau _{\rm min}}^{\tau _{\rm max}}d\tau \int \limits _0^{2\pi } d\phi_k 
\frac{{\cal G}(\tau,\phi_k)}{a_1+b\cos\phi_k}
=
2\pi \sqrt{\lambda _Y }L_S
{\cal G}(\tau _s,0)
\nonumber\\[1mm]
&\displaystyle
+
\int \limits_{\tau _{\rm min}}^{\tau _{\rm max}}d\tau \int \limits _0^{2\pi } d\phi_k 
\frac{{\cal G}(\tau,\phi_k)-{\cal G}(\tau _s,0)}{a_1+b\cos\phi_k},
\nn
&\displaystyle       
\int \limits_{\tau _{\rm min}}^{\tau _{\rm max}}d\tau \int \limits _0^{2\pi } d\phi_k 
\frac{{\cal G}(\tau,\phi_k)}{a_2+b\cos\phi_k}
=
2\pi \sqrt{\lambda _Y }L_X
{\cal G}(\tau _p,0)
\nonumber\\[1mm]
&\displaystyle
+
\int \limits_{\tau _{\rm min}}^{\tau _{\rm max}}d\tau \int \limits _0^{2\pi } d\phi_k 
\frac{{\cal G}(\tau,\phi_k)-{\cal G}(R ^\prime,\tau _p,0)}{a_2+b\cos\phi_k},
\label{int21}
\end{eqnarray}
where ${\cal G}(\tau,\phi_k)$
is a regular function of $\tau$ and $\phi_k$. 
The second terms in the right-hand side of these transformation do not include the leading terms and vanish in the our
approximation.

Following to  Eq.~(\ref{int21}) the quantities $\theta_{ij}^0$ from (\ref{srfin}) can be decomposed as
\ba
\theta_{ij}^0&=&\frac{\theta_{ij}^s}{a_1+b\cos \phi_k }+\frac{\theta_{ij}^p}{a_2+b\cos \phi_k }+\theta_{ij}^{rest},
\label{thsp}
\ea
where the quantities $\theta_{ij}^s$ and $\theta_{ij}^p$ contain the terms proportional to $1/kk_1$ and $1/kk_2$ and independent on the integration variables $\tau$ and $\phi_k$. They are obtained in the limit $m\to 0$, $\phi_k=0$ and $\tau\to \tau_s$ and $\tau\to  \tau_p$ for $s$- and $p$-peaks, respectively. The quantity $\mu$ at the peaks become $\mu \to\mu_s\equiv V_1/S$ and    
 $\mu\to \mu_p\equiv V_1/X$. The last term in Eq.~(\ref{thsp}), $\theta_{ij}^{rest}$, does not give the contribution to the leading approximation.

The quantities $\theta_{ij}^s$ and $\theta_{ij}^p$ are expressed in terms of respective Born $\theta ^B_i$ defined in Eq.~(\ref{thb}) 
\ba
\theta_{ij}^{s,p}=d_{j}^{s,p}\theta ^B_i,
\label{thsp1}
\ea
where
\ba
d_{1}^s=-4S,\;
d_{2}^s=4,\;
d_{3}^s=-\frac 2S,\;
\nn
d_{1}^p=-4X,\;
d_{2}^p=4,\;
d_{3}^p=-\frac 2X.
\label{thsp2}
\ea
Then the sums over $j$ can be explicitly calculated: 
\ba
\sum\limits_{j=1}^{3}\Biggl[\frac{R^\prime}{1+\tau_s-\mu_s}\Biggr]^{j-2}d^s_{j}
&=&-2
\frac{(S^\prime-R^\prime)^2+S^{\prime 2}}{S^\prime R^\prime},
\nonumber\\[2mm]
\frac{(1+\tau_s-\mu_s)}{R^\prime}d^s_{1}
&=&-
4
\frac{S^\prime }{R^\prime },
\nn[2mm]
\sum\limits_{j=1}^3\Biggl[\frac{R^\prime}{1+\tau_p-\mu_p}\Biggr]^{j-2}d^p_{j}
&=&-
2
\frac{(X^\prime+R^\prime)^2+X^{\prime 2}}{X^\prime R^\prime},
\nn[2mm]
\frac{(1+\tau_p-\mu_p)}{R^\prime}d^p_{1}
&=&-
4\frac{X^\prime}{R^\prime },
\label{ssum}
\ea
where we used $1+\tau_s-\mu_s=S^\prime/S$ and  $1+\tau_p-\mu_p=X^\prime/X$. 

Substitution of the decomposition (\ref{thsp}) into (\ref{srfin}) results in separation of $\sigma_R^F$  into two parts
corresponding to $s$- and $p$- collinear singularities. Integration over $R^\prime$ can be further  replaced by $z_1$ and $z_2$ for these two parts using the substitutions $R^\prime\to (1-z_1)S^\prime$ and $R^\prime \to (z_2^{-1}-1)X^\prime $:
\ba
\int\limits_0^{p_x^2-M_{th}^2}dR^\prime \to S^\prime \int\limits_{z_{1i}}^1 dz_1,
\nn  
\int\limits_0^{p_x^2-M_{th}^2}dR^\prime \to X^\prime \int\limits_{z_{2i}}^1 \frac {dz_2}{z_2^2},
\ea
where the lowest limits of integration over variables $z_{1,2}$ are defined by Eqs.~(\ref{z12m}). The expression from the r.h.s. of Eq.~(\ref{ssum}) are reduced as:
\ba
\frac{(S^\prime-R^\prime)^2+S^{\prime 2}}{S^\prime R^\prime}&=&\frac{1+z^2_1}{1-z_1},
\nn
\frac{(X^\prime+R^\prime)^2+X^{\prime 2}}{X^\prime R^\prime}&=&\frac{1+z^2_2}{z_2(1-z_2)}.
\ea

The obtained equations are combined as follows resulting in final expressions in the leading log approximation. The  substitution of Eq. (\ref{thsp}) into (\ref{srfin}) with dropped $\theta_{ij}^{rest}$ splits the expression $\sigma^F_R$ into two parts that correspond to $s$- and $p$-peaks 
according to the upper index in the first and second terms of $\theta_{ij}$ in the r.h.s. of Eq.~(\ref{thsp}): $\sigma^F_R=\sigma^F_s+\sigma^F_s$. Integration over $\tau$ and $\phi_k$ is performed using (\ref{int21}) in which the second terms in the r.h.s. have to be dropped. The arguments of  the structure functions ${\cal H}_i(Q^2+\tau R,\tilde x,\tilde z, \tilde p_t)$ are transferred into ${\cal H}_i(z_1Q^2,x_s,z_s,p_{ts})$ or ${\cal H}_i(z_2^{-1}Q^2,x_p,z_p,p_{tp})$ for $s$- or $p$-peaks respectively, where the quantities with the subscripts $s$ and $p$ are defined by Eq.~(\ref{shift}). Finally, the representation of $\theta^{s,p}_{ij}$ in  the form (\ref{thsp1},\ref{thsp2}) allows us to perform summation over $j$ as it was shown in Eqs.~(\ref{ssum}) and obtain the final expressions in the form:
\ba
\sigma_s^F&=&\frac{\alpha ^3S_x^2}{8  Mp_lS
}l_m
\int\limits_{z_{1i}}^{1}
dz_1  
\nn &&\times  
\sum\limits_{i=1}^4
\Biggl[
\frac{1+z_1^2}{1-z_1}\frac{ {\cal H}_i(z_1Q^2, x_s, z_s, p_{ts})\theta^B_{i}}{z_1^2Q^4}
\nn &&
-
2\frac{\theta^B_{i}{\cal H}_i(Q^2,x,z,p_t)}{(1-z_1)Q^4}
\Biggr],
\nn
\sigma_p^F&=&\frac{\alpha ^3S_x^2}{8  Mp_lS
}l_m
\int\limits_{z_{2i}}^{1}
\frac{dz_2}{z_2^2}  
\nn &&\times  
\sum\limits_{i=1}^4
\Biggl[
\frac{1+z_2^2}{z_2(1-z_2)}\frac{ z_2^2{\cal H}_i(z_2^{-1}Q^2, x_p, z_p, p_{tp})\theta^B_{i}}{Q^4}
\nn &&
-
2\frac{\theta^B_{i}{\cal H}_i(Q^2,x,z,p_t)}{(1-z_2)Q^4}
\Biggr].
\label{srfinll}
\ea
 Using Eq.~(\ref{born2}) we represent the products of $\theta^B_{i}$ and ${\cal H}_i$ 
through the Born cross section in the shifted kinematics for $s$- and $p$-peaks respectively:
\ba
&\displaystyle
\frac{ {\cal H}_i(z_1Q^2, x_s, z_s, p_{ts})\theta^B_{i}}{z_1^2Q^4}\qquad\qquad
\nn
&\displaystyle
=\frac{4M^2p_{ls}S}{\pi\alpha^2(z_1S-X)^2}\sigma_B(z_1S,z_1Q^2,x_s,z_s,p_{ts},\cos \phi_{hs}),
\nn
&\displaystyle
\frac{ z_2^2{\cal H}_i(z_2^{-2}Q^2, x_p, z_p, p_{tp})\theta^B_{i}}{Q^4}
\nn
&\displaystyle
=\frac{4M^2p_{lp}S}{\pi\alpha^2(S-z_2^{-1}X)^2}
\sigma_B(S,z_2^{-1} Q^2,x_p,z_p,p_{tp},\cos \phi_{hp}).
\nn
\ea
and obtain the expressions (\ref{sfsp}) and, after cancellation of the infrared divergence, (\ref{srvll}). Finally, the expressions in the leading log approximation, (\ref{sll1}), can be  obtained from (\ref{srvll}) using the integral representation for $\delta_{VR}$ that is defined by Eq.~(\ref{dvrll}). Indeed, the difference between (\ref{sfsp}) and (\ref{sll1}) is exactly  $\alpha/\pi\; \delta_{VR} \sigma^B$:
\ba 
\delta_{VR}&=&\frac{l_m}{2}\Biggl[\;
\int\limits_{z_{1i}}^1dz_1(1+z_1)-\int\limits_0^{z_{1i}}dz_1\frac{1+z_1^2}{1-z_1} 
\nn &&
+\int\limits_{z_{2i}}^1dz_2\frac{2+z_2+z_2^2}{z_2}-\int\limits_0^{z_{2i}}dz_2\frac{1+z_1^2}{1-z_1}
\Biggr].
\ea

\subsection{Electron structure function method}
Up to now the lowest order RC to SIDIS in the leading approximation have been considered. The second order RC to the cross section of unpolarized inclusive DIS withing leading order  was firstly estimated by Kripfganz, Mohring and Spiesberger in \cite{Spies} and were generalized to polarized DIS by our group \cite{AISh1998}. The approach to summing up  the leading logarithmic RC of all orders over $\alpha$ that involves the electron structure function was suggested by Fadin, Merenkov and Kuraev in \cite{kuraev1,kuraev2}. This method was applied for polarized inclusive DIS in \cite{AAM2004}. The main features of the method of the electron structure functions as well as the detailed comparison between different approaches for calculation of RC to polarized inclusive DIS is presented in
\cite{ESFRAD}.  

\begin{figure}[t!]\centering
\includegraphics[width=8.5cm,height=8.5cm]{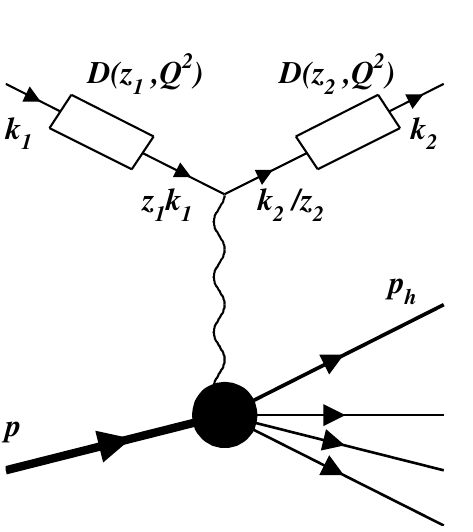}
\caption{The diagram for the cross section in the methods of the electron structure functions}
\label{esf}
\end{figure}

The cross section of SIDIS within the method of the electron structure functions (illustrated in Fig.~\ref{esf}) reads:
\ba
&\displaystyle
\sigma^{in}_{hL}=
\frac{S_x^2}{p_l}
\int\limits_{z_{1i}}^1dz_1D(z_1,Q^2)\int\limits_{{\hat z}_{2i}}^1\frac{dz_2}{z^2_2}D(z_2,Q^2)
\nonumber\\
&\displaystyle
\times
r^2\left(\frac{z_1}{z_2}Q^2\right)
\frac{\hat p_l
\sigma_{hard}(z_1S,z_1z_2^{-1}Q^2,\hat x, \hat z, \hat p_t,\cos \hat \phi_h)
}{(z_1S-X/z_2)^2}
,
\nn
\label{inll}
\ea
where $z_{1i}$ is defined by Eq.~(\ref{z12m}),  and
\ba
\hat z_{2i}=\biggl[1+\frac{p_x^2-(1-z_1)S^\prime-M_{th}^2}{X-V_2+z_1Q^2}\biggr]^{-1}.
\label{hz2}
\ea
The electron structure function $D(z_{1,2},Q^2)$ contains three terms 
\ba
D=D^\gamma+D^{e^+e^-}_N+D^{e^+e^-}_S,
\label{esf3}
\ea
where $D^\gamma$ describes the contribution of photon radiation, and $D^{e^+e^-}_N$ and $D^{e^+e^-}_S$ correspond to the process of the electron pair production  in nonsinglet
(by the single photon mechanism) and singlet (by the double
photon mechanism) channels, respectively \cite{kuraev1,kuraev2,AAM2004,ESFRAD}.
The explicit expressions for the components of ESF $D(z_{1,2},Q^2)$ are presented in Eqs.~(5-7) of  \cite{AAM2004}. 
 The coefficient $r^2$ in the integrand in Eq.~(\ref{inll}) results from ressumation of the vacuum polarization by leptons (\ref{dvac}),
\ba
r(Q^2)=\sum\limits_{i=0}^\infty\left(\frac{\alpha}{2\pi}\delta_{\rm vac}^l( Q^2)\right)^i= \biggl[1-\frac{\alpha}{2\pi}\delta_{\rm vac}^l( Q^2)\biggr]^{-1},
\ea
and represented in the form of the running coupling constant.

The hard cross section, $\sigma_{hard}$, in (\ref{inll}) is the radiative corrected SIDIS cross section excluding the leading log term \cite{AAM2004},
\ba
\sigma_{hard}=\sigma_B
+\sigma^{in}-\sigma^{in}_{1L}.
\ea
Here $\sigma^{in}$ and $\sigma^{in}_{1L}$ are defined by Eqs.~(\ref{srv}) and (\ref{sll1}) respectively. 
This cross section is  generalized to all orders of $\alpha$ as
\ba
\sigma_{hard}=\sigma_B+\sum\limits_{i=1}^\infty\alpha^i\sum\limits_{j=0}^{i-1}C_{ij}l_m^j+{\cal O}\left(\frac {m^2}{Q^2}\right),
\label{generalhard}
\ea
where the coefficients $C_{ij}$ do not depend on the electron mass and  are responsible for subleading contributions in each order of $\alpha ^i$. The formula (\ref{inll}) with $\sigma_{hard}$ given by Eq.(\ref{generalhard}) is the best approximation of RC from continuous spectrum (i.e., excluding the exclusive radiative tail) in SiDIS processes. 


 
The expressions for the shifted variables in (\ref{inll}) represent extension of such variables defined by Eqs.~(\ref{shift}):
\ba
&\displaystyle
\hat x=\frac{z_1Q^2}{z_1z_2S-X},\;
\hat z=\frac{zS_x}{z_1S-z_2^{-1}X},\;
\nonumber\\
&\displaystyle
\hat \lambda_Y=(z_1S-z_2^{-1}X)^2+4z_1z_2^{-1}M^2 Q^2,
\nonumber\\
&\displaystyle
\hat p_l=\frac{z S_x(z_1S-z_2^{-1}X)-2M^2(z_1V_1-z_2^{-1}V_2)}{2M\sqrt{\hat \lambda_Y}},\;
\nonumber\\
&\displaystyle
\hat p_t^2=\frac{z^2S_x^2}{4M^2}-\hat p_l^2-m_h^2,
\nonumber\\
&\displaystyle
\cos \hat \phi_h=\frac{z_2}{4z_1\hat p_t\sqrt{\hat \lambda_Y\lambda_1}}\Biggl[\biggl(z_1S+\frac X{z_2}\biggr)\biggl(2\frac{z_1}{z_2}zS_x \hat Q^2
\nonumber\\
&\displaystyle
+\biggl(z_1S-\frac X{z_2}\biggr)\biggl(z_1V_1-\frac{V_2}{z_2}\biggr)\biggr)-\hat \lambda_Y(z_1V_1+\frac{V_2}{z_2}\biggr)\Biggr].
\nn
\label{shift2}
\ea

We note that the expression for the cross section in the leading log approximation in Eq.~(\ref{sll1}) is reproduced from (\ref{inll}) by keeping the first order (non-trivial) terms in series over $ \alpha $ of the electron structure function \cite{JSW},
 $r(Q^2)$, and the $\sigma_{hard}$: 
\ba
D(z,Q^2)&\to &\delta(1-z)+\frac \alpha{2\pi}l_mP(z),
\nn
r(Q^2)&\to & 1+\frac \alpha{\pi}\delta^l_{\rm vac}(Q^2),
\nn
\sigma_{hard} & \to & \sigma_B.
\ea

\section{Applying leading log result to exclusive radiative tail}
\label{ert}

Exact contribution calculated in \cite{AI2019} 
reads:
\ba
&\displaystyle
\sigma^{ex}_{R}=-\frac{\alpha^3SS_x^2}{2^8\pi^4Mp_l\lambda_S\sqrt{\lambda_Y}}
\int\limits_{\tau _{\rm min}}^{\tau _{\rm max}}d\tau
\int\limits_0^{2\pi}d\phi_k
\nonumber \\&\displaystyle\times
\sum_{i=1}^{4}
\sum_{j=1}^3
\frac{{\mathcal H}^{ex}_i(Q^2+\tau R_{ex},\tilde W^2_{ex},\tilde t_{ex})\theta^0_{ij}R_{ex}^{j-2}}{(1+\tau-\mu)(Q^2+\tau R_{ex})^2},
\label{sre1}
\ea
where
\ba
&\displaystyle
R_{ex}=\frac{p_x^2-m_u^2}{1+\tau-\mu},
\tilde W^2_{ex}=W^2-(1+\tau R_{ex}),
\nn
&\displaystyle
\tilde t_{ex}=t+R_{ex}(\mu-\tau).
\ea

The leading log terms can be extracted from $\theta^0_{ij}$ using methods of Subsection \ref{llext}. For these analyses it is necessary to keep in mind that only angular integrations, i.e. over $\tau $ and $\phi_k$, have to be performed 
for the exclusive tail.
However in the present Section we will used result of Eqs.~(\ref{dsrsp1}) and (\ref{inll}).        

Similar to  SiDIS the Born cross section of the exclusive process,
\ba
l(k_1)+n(p)\rightarrow l(k_2)+h(p_h)+u(p_u)
\label{excprocess0}
\ea
($p_u^2=m_u^2$),  can be presented in the form
of convolution leptonic and hadronic tensors
(\ref{wl})
\ba
d\sigma^{ex} _B=\frac{(4\pi\alpha)^2}{2SQ^4}W^{ex}_{\mu \nu}(q,p,p_h)L^{\mu \nu}_B
d\Gamma^{ex} _B,
\label{wle}
\ea
$L^{\mu \nu}_B$ was defined earlier by Eq.~(\ref{ltborn}) whereas 
\ba
&\displaystyle
W^{ex}_{\mu\nu}(q,p,p_h)=\sum\limits_{i=1}^4 w^i_{\mu\nu}(q,p,p_h){\cal H}^{ex}_i
\label{ht1e}
\ea
and the quantities $w^i_{\mu\nu}$ have the same structure as in (\ref{ht1}).
As it was presented in Appendix A of \cite{AIO}, the exclusive structure functions ${\mathcal H}^{ex}_i$ can be expressed through the standard set of the two-fold cross sections $d\sigma_L/d\Omega$, $d\sigma_T/d\Omega$, $d\sigma_{LT}/d\Omega$ and $d\sigma_{TT}/d\Omega$. 

The phase space 
can be expressed through $d\Gamma_B$ (\ref{dgin}) as
\ba
d\Gamma_B^{ex}&=&
d\Gamma _B\frac{d^3p_u}{(2\pi)^32p_{u0}}\delta^4(p+k_1-k_2-p_h-p_u).
\nn
\label{dgex}
\ea

As a result
\ba
&\displaystyle
\frac{d\sigma^{ex} _B}{dxdydzdp^2_td\phi_h}=\frac 1{(2\pi)^3}\delta\Biggl(\frac{\sqrt{\lambda_Y}p_l}{M}+W^2+m_h^2-m_u^2
\nn 
&\displaystyle
-\frac{z S_x(S_x+2M^2)}{2M^2}\Biggr)
\frac{\pi \alpha^2 S^2_x}{4MQ^4p_lS}
\sum\limits_{i=1}^4\theta^B_i{\cal H}^{ex}_i
\nn
&\displaystyle
=\delta(z_0-z)
\bar \sigma^{ex}_B(S,Q^2,x,p_t^2,\cos \phi_h),
\label{exc5}
\ea
where
\ba
z_0=\frac{2M(\sqrt{\lambda_Y}p_l+M(W^2+m_h^2-m_u^2))}{S_x(S_x+2M^2)}.
\ea 
Here and below the symbol $\bar \sigma^{ex}$ is used to denote the four-fold cross section of exclusive processes, and the original symbol $\sigma^{ex}$ is kept to represent the five-fold contribution of exclusive processes to RC to the SiDIS cross section.

After tensor convolution  and integration over $z$ using $\delta$-function the Born cross section of the exclusive process
reads
\ba
&\displaystyle
\bar \sigma^{ex}_B(S,Q^2,x,p_t,\cos \phi_h)\equiv
\frac{d\sigma^{ex}_B}{dxdydp_t^2d\phi_h}
\nonumber \\ 
&\displaystyle
=\frac{\alpha^2 M S_x}{16\pi^2Q^4Sp_l(S_x+2M^2)}
\sum\limits_{i=1}^4{\cal H}_i^{ex}(Q^2,W^2,t)
\theta_i^B.
\label{exc6}
\ea
Here for exclusive process
\ba
&&
\; p_t^2=p_{h0}^2-p_{l}^2-m_h^2,
\nn
&&
\; p_l=\frac{S_x(W^2+m_h^2-m_u^2)-2V_-(S_x+2M^2)}{2M\sqrt{\lambda_Y}},
\nn
&&
p_{h0}=\frac{W^2+m_h^2-m_u^2-2V_-}{2M}.
\label{ptex}
\ea

The general leading log formulae are given by expressions (\ref{dsrsp1}). These formulae are applicable for the contribution of the exclusive radiative tail to SiDIS process in which the five-fold cross section (\ref{exc5}) has to be used for $\sigma_B$ in (\ref{dsrsp1}). However, the cross section (\ref{exc5}) contains the $\delta $-function because of the fixed mass of the unobserved hadronic state. This $\delta $-function is used to integrate over $z_1$ or $z_2$, so the final expressions do not contain the integration as in (\ref{sll1}) and expressed in terms of the four-fold born cross section of exclusive process (\ref{exc6}). We demonstrate the derivation of the leading log formulae for exclusive radiative tail by obtaining the formulae for the cross section (\ref{exc5}) in $s$- and $p$-peaks (i.e., in the shifted kinematics) and analytic integration using the $\delta $-function. 

 The phase space in the shifted kinematics for $s$- and $p$- peaks is:
%
\ba
d\Gamma_B^{ex\; s}&=&
d\Gamma^s _B\frac{d^3p_u}{(2\pi)^32p_{u0}}\delta^4(p+z_1k_1-k_2-p_h-p_u)
\nn
&=&\frac 1{(2\pi)^3}d\Gamma^s _B
\delta(
M^2+m_h^2-m_u^2-z S_x
\nn &&
+z_1S^\prime+V_2-X),
\nn
d\Gamma_B^{ex\; p}&=&
d\Gamma^p _B\frac{d^3p_u}{(2\pi)^32p_{u0}}\delta^4(p+k_1-k_2/z_2-p_h-p_u)
\nn
&=&\frac 1{(2\pi)^3}d\Gamma^s _B
\delta(
M^2+m_h^2-m_u^2-z S_x
\nn &&
+S-V_1-X/z_2),
\label{dgexsp}
\ea
where $d\Gamma^{s,p} _B$ are the phase spaces for SiDIS Born processes in the shifted kinematics.

As a result
\ba
&\displaystyle
\frac{d\sigma^{ex\; s} _B}{dx_sdy_sdz_sdp^2_{ts}d\phi_{hs}}=\frac 1{(2\pi)^3}
\delta(
M^2+m_h^2-m_u^2-z S_x
\nn
&\displaystyle
+z_1S^\prime
+V_2-X)\frac{\pi \alpha^2 (z_1S-X)^2}{4z_1^3MQ^4p_{ls}S}
\sum\limits_{i=1}^4\theta^B_{i\;s}{\cal H}^{ex}_{i\;s},
\nn
&\displaystyle
\frac{d\sigma^{ex\; p} _B}{dx_pdy_pdz_pdp^2_{tp}d\phi_{hp}}=\frac 1{(2\pi)^3}
\delta(
M^2+m_h^2-m_u^2-z S_x
\nn 
&\displaystyle
+S-V_1-X/z_2)
\frac{\pi \alpha^2 z_2(S-X/z_2)^2}{4MQ^4p_{lx}S}
\sum\limits_{i=1}^4\theta^{B}_{i\;p}{\cal H}^{ex}_{i\;p}.
\nn 
\label{sllex10}
\ea

After substitution (\ref{sllex10}) into Eqs.~(\ref{dsrsp1}) taking into account
that for the exclusive process $M_{th}^2$ incoming into (\ref{z12m}) is equal 
to the undetected hadron mass square $m_u^2$ 
after integration over $z_{1,2}$
we can obtain that 
\ba
\sigma^{ex}_{1L}=\sigma^{ex\;s}_{1L}+\sigma^{ex\;p}_{1L},
\label{sllex1}
\ea
where
\ba
&\displaystyle
\sigma^{ex\;s}_{1L}=\frac{\alpha}{2\pi}l_m
\frac{1+z_{1e}^2}{1-z_{1e}}
\frac{p_{lse}}{p_l}\frac{S^2_x}{S^\prime }\left[\frac 1{z_{1e}S-X}+\frac 1{2M^2}\right]
\nonumber\\[2mm]&\displaystyle\times
\bar \sigma^{ex}_B(z_{1e}S,z_{1e}Q^2,x_{se},p_{tse},\cos \phi_{hse}),
\nonumber\\[2mm]
&\displaystyle
\sigma^{ex\;p}_{1L}=\frac{\alpha}{2\pi}l_m
\frac{1+z_{2e}^2}{1-z_{2e}}
\frac{p_{lpe}}{p_l}\frac{S^2_x}{X^\prime }\left[\frac 1{S-X/z_{2e}}+\frac 1{2M^2}\right]
\nonumber\\[2mm]&\displaystyle\times
\bar \sigma^{ex}_B(S,Q^2/z_{2e},x_{pe},p_{tpe},\cos \phi_{hpe}),
\label{sexll1}
\ea
and  the quantities $z_{1,2e}$ are defined using Eq.~(\ref{z12m}) as
\ba
z_{1,2e}&=&z_{1,2i}(M_{th}^2\to m_u^2).
\label{z12me}
\ea
The variables with the indexes $se$ ($pe$)
are calculated from Eqs.~(\ref{shift}) using the replacements: i)  $z_1\to z_{1e}$ and
$z \to (M^2+m_h^2-m_u^2+z_{1e}S^\prime+V_2-X)/S_x$ for $\sigma^{ex\;s}_{1L}$, and ii) $z_2\to z_{2e}$ and
$z \to (M^2+m_h^2-m_u^2+X^\prime/z_{2e}+S-V_1)/S_x$ for $\sigma^{ex\;p}_{1L}$.

The generalization on the high orders is performed similarly to (\ref{inll}). This formula is applicable to the case of exclusive radiative tail, however again, the $\sigma_{hard}$ in (\ref{inll}) have to be presented through the product of the four-fold cross section and respective $\delta$-function. This $\delta$-function is then used to integrate over one of two integration variables $z_1$ or $z_2$.
\ba
d\hat \Gamma_B^{ex}&=&
d\hat \Gamma _B\frac{d^3p_u}{(2\pi)^32p_{u0}}\delta^4(p+z_1k_1-k_2/z_2-p_h-p_u)
\nn
&=&\frac 1{(2\pi)^3}d\hat \Gamma _B\delta(
M^2+m_h^2-m_u^2-z S_x
\nn &&
-z_1z_2^{-1}Q^2+z_1(S-V_1)+z_2^{-1}(V_2-X)).
\label{dgexh}
\ea
we can used (\ref{inll}) with $M_{th}^2=m_u^2$. As a result
\ba
&\displaystyle
\sigma^{ex}_{hL}=\frac{S_x^2}{p_l}
\int\limits_{z_{1e}}^1dz_1D(z_1,Q^2)\int\limits_{{\hat z}_{2e}}^1\frac{dz_2}{z^2_2}D(z_2,Q^2)
\nonumber\\
&\displaystyle
\times
r^2\left(\frac{z_1}{z_2}Q^2\right)
\hat p_l
\biggl[\frac 1{z_1S-X/z_2}+\frac 1{2M^2}\biggr]
\delta(
M^2+m_h^2
\nn 
&\displaystyle
-m_u^2-z S_x
-z_1z_2^{-1}Q^2+z_1(S-V_1)+z_2^{-1}(V_2-X))
\nonumber\\
&\displaystyle
\times
\bar \sigma^{ex}_{B}(z_1S,\frac{z_1}{z_2}Q^2,\hat x, \hat p_t,\cos \hat \phi_h),
\label{exllz1z2}
\ea
and the shifted quantities are defined by Eq.~(\ref{shift2}).

\begin{figure}[t]\centering
\scalebox{0.3}{\includegraphics{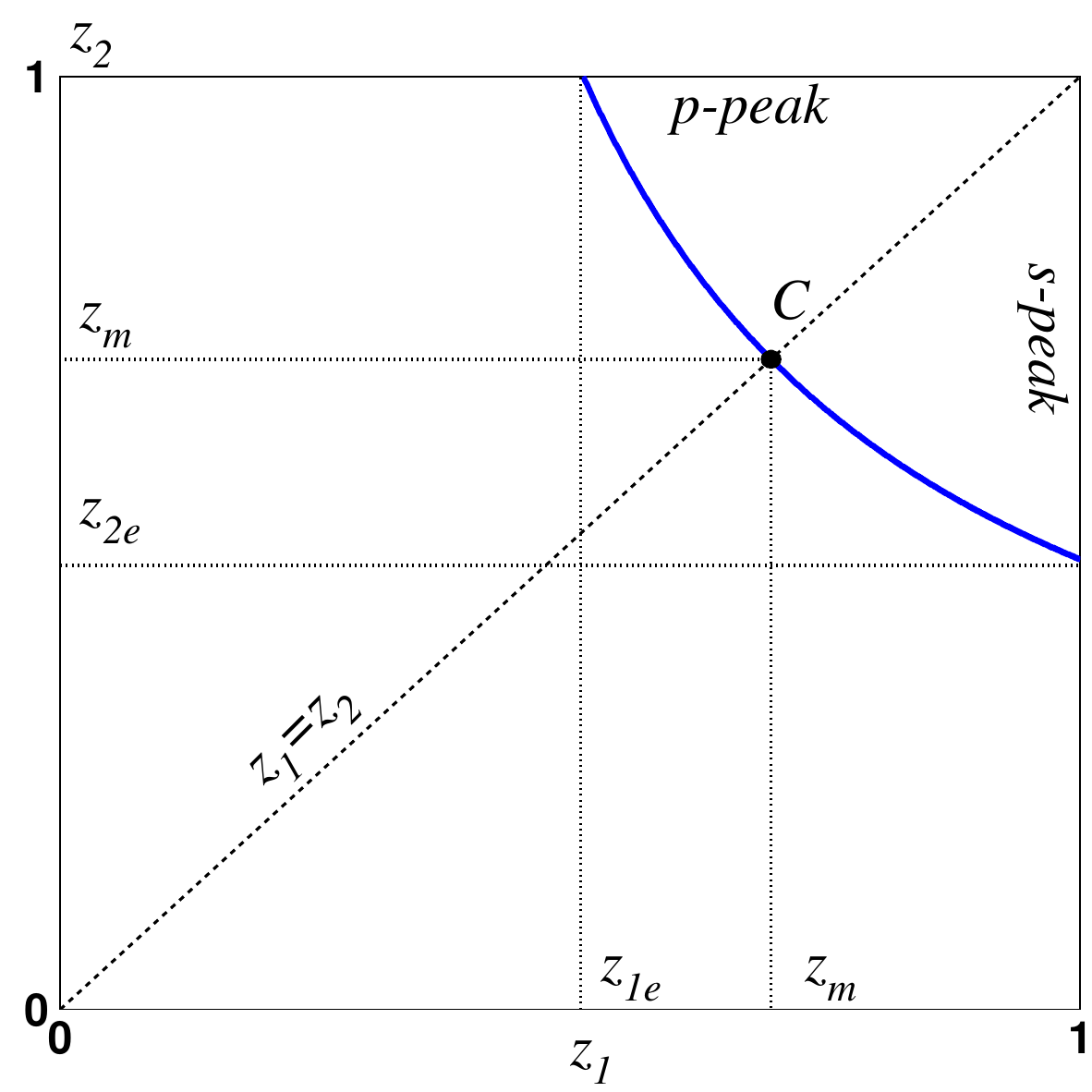}}
\caption{Integration area (blue line) in the plane $(z_1,z_2)$.
The quantities $z_{1,2e}$ and $z_m$ are defined by Eqs.~(\ref{z12m}), (\ref{z12me}) and (\ref{zmmm}).}
\label{z1z2cor}
\end{figure}

Due to the functional relationship between $z_1$ and $z_2$ induced by $\delta$-function, the integration area in (\ref{exllz1z2}) is the solid curve shown in Figure \ref{z1z2cor}. The integrand has two sharp peaks in the areas when $z_1$ and $z_2$ close to $1$ which come from the functions $D(z_1,Q^2)$ and $D(z_2,Q^2)$ and interpreted as $s$- and $p$-peaks respectively. There are two ways to remove $\delta$-function in (\ref{exllz1z2}) performing integration either over $z_1$ or $z_2$, 
however 
it is natural to split integration region by the point C with coordinates ($z_m$,$z_m$) (crossing the integration area and the line $z_1=z_2$; Figure \ref{z1z2cor}), thus isolating $s$- and $p$-peak  peaks in separate contributions,  
\ba
&\displaystyle
z_m=\frac1{2(S-V_1)}\biggl[
S_x+m^2_u-p_x^2-2V_-
\nn 
&\displaystyle
+\sqrt{(p_x^2-S_x-m^2_u+2V_-)^2+4(S-V_1)(X-V_2)}\biggr].
\nn
\label{zmmm}
\ea  
The result of this integration is:
\ba
&\displaystyle
\sigma^{ex}_{hL}=\frac{S_x^2}{p_l}
\int\limits_{z_m}^1
\Biggl[dz_1\frac{D(z_1,Q^2)D(\hat z_{2},Q^2)}{X-V_2+z_1Q^2}
r^2\left(\frac{z_1}{\hat z_{2}}Q^2\right)
\nn
&\displaystyle
\times
\hat p_{l2}
\biggl[\frac 1{z_1S-X/\hat z_{2}}+\frac 1{2M^2}\biggr]
\nn
&\displaystyle
\times
\bar \sigma^{ex}_{hard}(z_1S,\frac{z_1}{\hat z_{2}}Q^2,\hat x_2, \hat p_{t2},\cos \hat \phi_{h2})
\nn
&\displaystyle
+
\frac{dz_2}{z_2^2}\frac{D(\hat z_{1},Q^2)D(z_2,Q^2)}{S-V_1-Q^2/z_2}
r^2\left(\frac{\hat z_{1}}{z_{2}}Q^2\right)
\nn
&\displaystyle
\times
\hat p_{l1}
\biggl[\frac 1{\hat z_{1}S-X/z_{2}}+\frac 1{2M^2}\biggr]
\nn 
&\displaystyle
\times
\bar \sigma^{ex}_{hard}(\hat z_{1}S,\frac{\hat z_{1}}{z_{2}}Q^2,\hat x_1, \hat p_{t1},\cos \hat \phi_{h1})
\Biggr]
\label{exll}
\ea
where
\ba
&\displaystyle
\hat z_{2}=\biggl[1+\frac{p_x^2-(1-z_1)S^\prime-m_u^2}{X-V_2+z_1Q^2}\biggr]^{-1},
\nn
&\displaystyle
\hat z_{1}=1-\frac{p_x^2+(1-1/z_2)X^\prime-m_u^2}{S-V_1-Q^2/z_2}.
\label{hz2e}
\ea
The shifted variables in (\ref{exll}) are defined by Eqs.~(\ref{shift2}) with the transformations i) $z_2\to \hat z_2$ and $z\to (M^2+m_h^2-m_u^2+z_1(S-V_1)+(X-V_2)/\hat z_2)/S_x$ for the integrand over $z_1$ in (\ref{exll}), and ii) $z_1\to \hat z_1$ and $z\to (M^2+m_h^2-m_u^2+\hat z_1(S-V_1)+(X-V_2)/z_2)/S_x$ for the integrand over $z_2$. 

The cross section $\bar \sigma^{ex}_{hard}$ in (\ref{exll}) is the cross section of exclusive process with the lowest order RC excluding the leading log terms. The use of this cross section in (\ref{exll}) instead of $\sigma^{ex}_B$ allows to account for subleading effects in the RC of higher order. Formally,  
\ba
\bar \sigma^{ex}_{hard}=\bar \sigma^{ex}_{B}+\bar \sigma^{ex}_{RC}-\bar \sigma^{ex}_{1L},
\ea    
where $\bar \sigma^{ex}_{RC}$ is the cross section of exclusive processes with the lowest order RC 
that is given by Eq.~(55) of \cite{EXCLURAD}. We rewrote this cross section in terms of the variables that are used in this analysis, i.e., $d\sigma/(dW^2dQ^2d\Omega_h) \to d\sigma/(dxdydp_t^2d\phi_h)$:
\ba
\bar
\sigma^{ex}_{RC}&=&-\frac{\alpha ^3MS_x}{2^8\pi^4(S_x+2M^2)S Q^2}\int\limits_0^{v_m}dv\int \frac{d\Omega_k}f
\nn &&\times \sum\limits_{i=1}^4\biggl[
\frac{\theta_i {\cal H}_i^{ex}}{\tilde Q^4(p_l-v/2M\sqrt{\lambda_Y})}
-4F_{IR}\frac{\theta_i^0 {\cal H}_i^{ex\;0}}{\tilde Q^4p_l}\biggr]
\nn
\nn &&+\frac\alpha\pi(\delta^{ex}_{VR}+\delta_{\rm vac})\bar \sigma^{ex}_{B},
\ea 
where
\ba
\delta^{ex}_{VR}&=&(l_m-1)\log\frac {v_m^2}{S^\prime X^\prime}
+\frac 32l_m-2
\nn &&
-\frac 12 \log^2 \frac {S^\prime}{X^\prime}
+{\rm Li}_2\biggl[1-\frac{M^2Q^2}{S^\prime X^\prime}\biggr]-\frac {\pi ^2}6,
\nn
v_m&=&W^2+m_h^2-m_u^2-\frac{S_x+2M^2}{M} m_h.
\ea

Similarly, the leading order terms are obtained from (58) of \cite{EXCLURAD} and have the form in terms of the variables $x$, $y$, $p_t^2$, and $\phi_h$: 
\ba
&\displaystyle
\bar \sigma^{ex}_{1L}=\frac\alpha\pi\delta_{\rm vac}^l(Q^2)\bar\sigma^{ex}_B(S,Q^2,x,p_t,\cos \phi_h)
\nonumber\\[2mm]&\displaystyle
+\frac{\alpha S^2_x}{2\pi p_lS^\prime }l_m\int\limits_{z_{1m}}^1dz_1
P(z_1)
\left[\frac 1{z_{1}S-X}+\frac 1{2M^2}\right]
\nonumber\\[2mm]&\displaystyle\times
p_{ls}
\bar\sigma^{ex}_B(z_{1}S,z_{1}Q^2,x_{s},p_{ts},\cos \phi_{hs})
\nonumber\\[2mm]&\displaystyle
+\frac{\alpha S^2_x}{2\pi p_lX^\prime}l_m
\int\limits_{z_{2m}}^1\frac{dz_2}{z_2^2}
P(z_2)
\left[\frac 1{S-X/z_{2}}+\frac 1{2M^2}\right]
\nonumber\\[2mm]&\displaystyle\times
p_{lp}
\bar\sigma^{ex}_B(S,Q^2/z_{2},x_{p},p_{tp},\cos \phi_{hp}),
\label{sexll1}
\ea
where $\delta_{\rm vac}^l(Q^2)$ is defined by Eq.~(\ref{dvac}),  the lowest limits of integration has a form:
\ba
z_{1m}&=&1-v_m/S^\prime,\nn
z_{2m}&=&(1+v_m/X^\prime)^{-1},
\label{z12m1}
\ea
and the splitting function $P(z)$ is defined by Eqs.~(\ref{spf},\ref{spf1}).

\section{Numerical  results} \label{SectNum}
\begin{figure*}[t]\centering
\scalebox{0.85}{\includegraphics{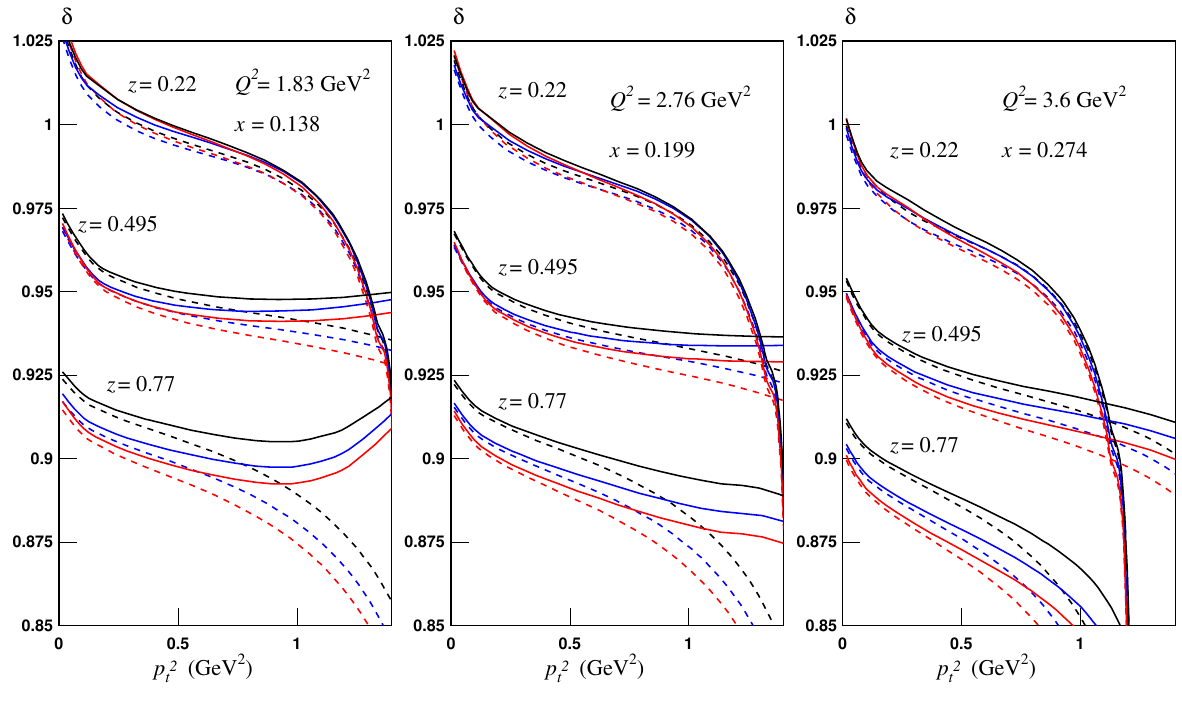}}
\caption{The $p_t$-dependence of RC factor for the cross sections of semi-inclusive $\pi ^+$ electroproduction averaged over $\phi_h$ with the lepton beam energy equal 10.65 GeV. Solid lines show the total RC factor, and 
dashed lines represent RC factor calculated excluding the exclusive radiative tail. The blue, red and black lines show RC factor calculated exactly, using the methods of the  leading log approximation in the lowest order in respect of $\alpha$, and the method of the electron structure functions, respectively. }
\label{fgn1}
\end{figure*}

\begin{figure*}[t]\centering
\scalebox{0.85}{\includegraphics{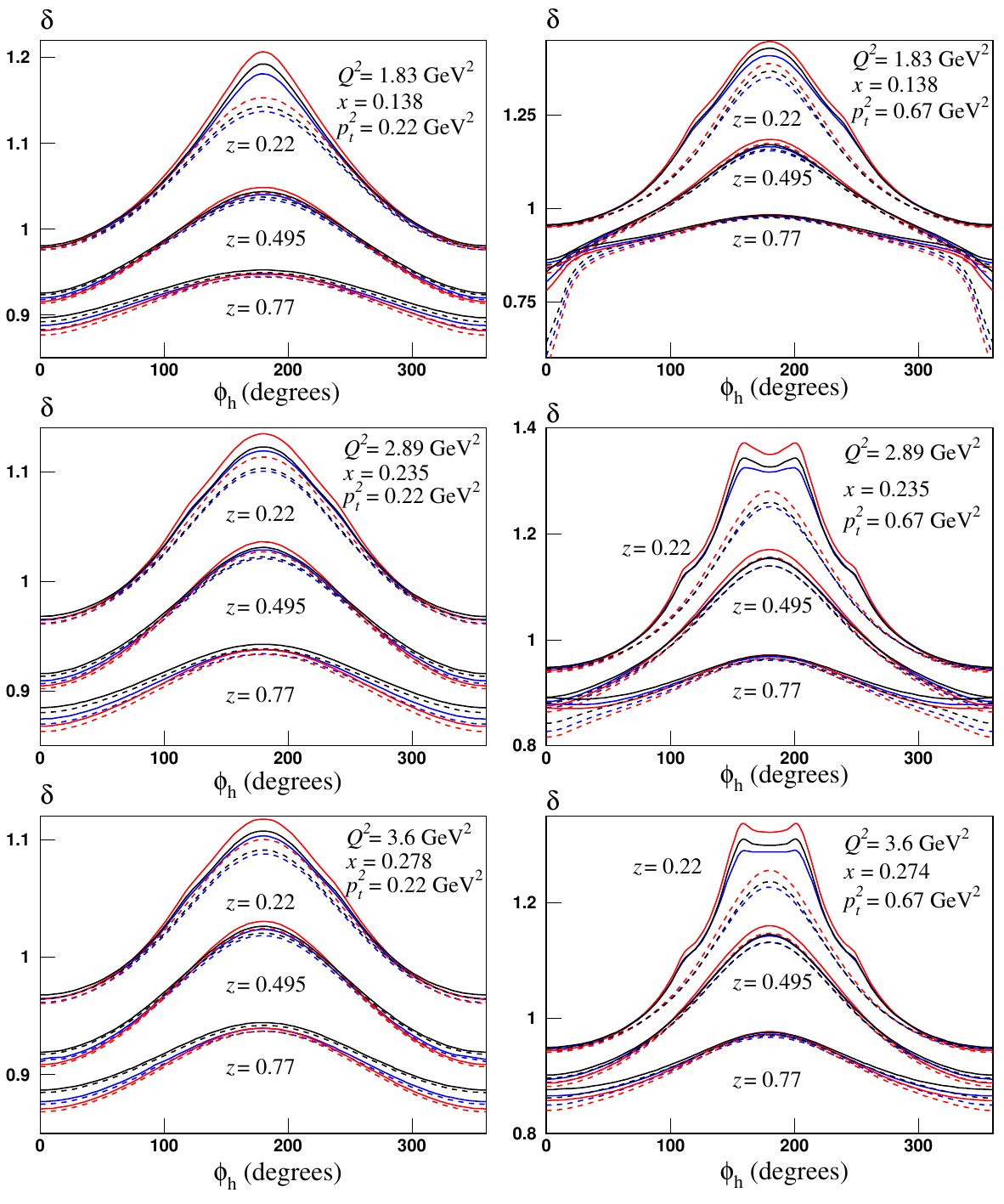}}
\caption{The $\phi_h$-dependence of RC factor for semi-inclusive $\pi ^+$ electroproduction for the lepton beam energy equal 10.65 GeV. Solid lines show the total RC factor, and 
dashed lines represent RC factor calculated excluding the exclusive radiative tail. The blue, red and black lines show RC factor calculated exactly, using the methods of the  leading log approximation in the lowest order in respect of $\alpha$, and the method of the electron structure functions, respectively. }
\label{fgn2}
\end{figure*}
The main characteristics used in the RC procedure of experimental data analysis  is the RC factor defined as a ratio of radiative corrected cross section to the Born contribution
\ba
\delta=\frac{\sigma_{obs}}{\sigma_B}.
\label{RCfactor}
\ea
For numerical estimates we applied the parametrization of the  SIDIS structure functions in Wandzura-Wilczek-type approximation \cite{WWSIDIS}. The exclusive structure functions are expressed through the two-fold cross sections $d\sigma_L/d\Omega$, $d\sigma_T/d\Omega$, $d\sigma_{LT}/d\Omega$ and $d\sigma_{TT}/d\Omega$ using MAID2007 parametrization \cite{MAID2007}. The $p_t$-dependence of RC factor, $\delta$, constructed from the  Born and observed cross sections of semi-inclusive $\pi ^+$ electroproduction averaged over $\phi_h$ is presented in Fig.~\ref{fgn1}. The solid lines show the total correction, dashed lines represent the correction excluding the exclusive radiative tail. The difference between exact and leading RC increases with growing $z$ and $p_t$. The $\phi_h$-dependence of  RC factor constructed from completely differential cross sections are presented in Fig.~\ref{fgn2}.  The RC factor reaches its maximum value at the region near $\phi_h=180^o$ and small $z$. In certain cases the curves for the RC factor are not smooth, e.g., for angles $\phi_h=160/200^o$ and $\phi_h=110/260^o$ in the right column plots in Fig.~\ref{fgn2}. This reflects the contributions of the exclusive radiative tail that is not small in these kinematic regions. The RC factor can be both higher and lower than one. The calculated (observed) RC factor is always a trade-off between i) the exclusive radiative tail contribution that is always positive, ii) semi-inclusive RC that can be negative because of the contribution of the vertex function, and iii) the vacuum polarization contribution that is always positive.
The blue lines in Figs.~\ref{fgn1} and \ref{fgn2} represent the RC factors calculated using the exact equations (\ref{srv}) and (\ref{sre1}). The red and black curves show the RC factor in the leading log 
approximation in the lowest order (\ref{sll1},\ref{sllex1}) and in all orders in respect to $\alpha$ (\ref{inll},\ref{exll}), respectively.
In all cases dashed and solid lines show the pure semi-inclusive RC and total RC, where the total RC additionally includes the contribution of the exclusive radiative tails.
When estimating the high order effects we restrict our consideration only to the leading orders terms, i.e., in (\ref{inll}) and (\ref{exll}) we put $\sigma_{hard}\equiv \sigma_B$ and $\bar \sigma_{hard}^{ex\; s,p}\equiv \bar \sigma_B^{ex}$  
respectively. The integrand in equations (\ref{inll}, \ref{exll}) contains the the cross section of hard photon radiation is the result of numeric multidimensional integration over the kinematics of the photon, so the implementation of eqs. (\ref{inll}, \ref{exll}) to our codes for numeric evaluation of RC would require a new level of results in the software development and will be a subject of a separate analysis.

For purposes of numerical analysis we had to modify the cross sections in (\ref{RCfactor}) to provide clear and well interpreted comparison of the exact and leading log RC. This is because the difference between exact and leading log RC comes not only because of difference in exact and leading log formulae for $\sigma_{in}$, what is one of the main focuses of the numerical analysis, but also due to a quite strong effect from the contribution from the vacuum polarization induced by $\mu$-, $\tau$-leptons and hadrons [see Fig.~\ref{fgrc}(b)] that are not included in the leading log formulae. The latter contribution [denote it as $\delta_{\rm vac}^{nlo}(Q^2)$]
 is trivial and is not of interest for the numerical comparison, so, we added  $\delta_{\rm vac}^{nlo}(Q^2)$ to $\delta_{\rm vac}^l(Q^2)$ in Eq.~(\ref{sll1}) and multiply the integrand in (\ref{inll}) on $1+\alpha \delta_{\rm vac}^{nlo}(z_1Q^2/z_2)/\pi$ to mask its effects in the difference between exact and leading log formulae.

\section{Discussion and Conclusion}\label{SectDiscConc}

In this paper  we presented the analytic expressions for RC to the SiDIS cross section derived analytically in the leading log approximation, that have a simple analytic form and were not explicitly presented and derived for SiDIS. We demonstrated three distinct approaches allowing for derivation of the expressions based on different theoretical and computational approaches. The ways in which the results are derived, clarify a quite complicated structure of the  exact formulae in \cite{AI2019} and further convince theoreticians and experimentalists dealing with practical data analyses in modern SiDIS measurements in using these results in data analyses and Monte Carlo generators.   

Specifically, we calculated RC in leading log approximation using three different approaches. First we applied the standard approach in the leading log approximation \cite{DeRujula, Blumlein,Spies,AI2012,APsp,Dsp,GLsp} and calculated the RC from the scratch. In this approach the only terms contributing to the cross section in the leading log approximation are extracted and kept, i.e., the poles that correspond to radiation collinear to initial and final electron (i.e., the terms that contain $1/kk_1$ and $1/kk_2$ and do not include the electron mass in the numerator). Integration over the photon angles can be performed analytically. Then, all these terms are combined resulting in the factorized form traditional for leading log calculations, i.e., the Born cross section in the so-called shifted kinematics depicted in Fig.~\ref{vect} for which the 3-vector of the virtual photon is shifted in the plane XOZ and the angle of this shift is determined by the photon energy (or equivalent variable $z_{1,2}$ for the photon emitted  collinear to the initial or final electron line), so there is a remaining one-dimensional integration variable in the final leading log formulae. This calculation resulted in exact expression for the term $A$ in (\ref{lonlo}). The infrared divergence is canceled in the usual way so the final formula (\ref{sll1}) is infrared free. The second approach is based on explicit extraction of the leading log contribution from the exact formulae presented in \cite{AI2019} by collecting all terms contribution to RC  in the leading log approximation after integration over
photon angles, and combined them into the final expression exactly coinciding to the expression obtained in the first approach.
Third, we used
the method of the electron structure functions \cite{AAM2004,kuraev1,kuraev2}. In this approach the QED radiative corrections to the corresponding cross
sections can be written as a convolution of  the two electron
structure functions corresponding to multiple real photon emission along to the initial and final electron and the Born cross section with shifted kinematics. Traditionally, these RC include effects
caused by loop corrections and soft and hard collinear
radiation of photons and $e^+e^-$ pairs. This method can be improved by including effects due
to radiation of one noncollinear photon. The corresponding
procedure results in a modification of the hard part of the cross
section, which takes the lowest-order correction into account
exactly and allows going beyond the leading approximation \cite{ESFRAD}.

Recently, Liu et al. \cite{Liu:2021jfp}
proposed a QCD-like factorization  to take into account the QED RC
 to the experimentally measured cross
sections of both inclusive and semi-inclusive lepton-nucleon DIS. This approach is similar to the approach for RC calculations involving 
the formalism of electron structure functions \cite{kuraev1,kuraev2,ESFRAD,AAM2004}. Since this approach is one of the three approaches we used in this paper  
 the resulting formula in the leading approximation (\ref{inll}) has to be comparable to Eq.~(3.30) obtained by Liu et al. \cite{Liu:2021jfp}. We note however, that the comparison deserves some comments. 

First, the lowest limits of integration in \cite{Liu:2021jfp}, $\xi_{min}$ and $\zeta_{min}$, are given by Eqs.~(2.24a) and (2.24b) and identical for both inclusive and semi-inclusive RC. The expressions for $\xi_{min}$ and $\zeta_{min}$ are calculated ignoring the restriction of the photon phase space by the pion threshold. The formulae for the lower limits of integration $z_{1m}$ and $z_{2m}$, that are given by Eq.~(11) (and subsequent formula) of \cite{AAM2004} for inclusive case and Eqs.~(\ref{z12m},\ref{hz2}) in the present paper. The formulae for $z_{1m}$ and $z_{2m}$ are not identical for DIS and SIDIS. 
This is expected because they can depend on $x$ and $y$ for DIS case, and on all five variables (\ref{setvar}) that describe the kinematics of SIDIS process. 
These formulae for inclusive case contain the term $z_{th}$  and reproduce $\xi_{min}$ and $\zeta_{min}$ when this term tends to zero.

 We believe that the pion threshold is necessary for both DIS and SIDIS RC to appropriately separate the contributions of the parts of the total RC with a single hadron and a continuum of particles in the final unobserved hadronic state. 
These two types of the contributions to RC require different models of hadronic structure (e.g., DIS/SIDIS structure functions for continuum of unobserved particles and form factors for the elastic radiative tail or exclusive structure functions for the exclusive radiative tail) and different models for the phase of space of unobserved particles (fixed invariant mass of the final hadronic state reduces the number of integration over the photon kinematical variables by one).
Ignoring the pion threshold in the formula for RC implies that the elastic radiative tail for DIS RC and exclusive RC for SIDIS RC can be obtained from the expressions for RC for continuum of particles by their extrapolation through the pion threshold and applying the obtained approximate formulae for the RC with one hadron in the final unobserved hadronic state. This approximation is poor and is not used since the seminal paper of Mo and Tsai \cite{Mo-Tsai} for calculation of RC in DIS.

Second, the expressions for the electron structure functions that are constructed and used in the formalism of the electron structure functions \cite{kuraev1,kuraev2,ESFRAD,AAM2004} are not identical to the lepton distribution and lepton fragmentation functions obtained by Liu et al. \cite{Liu:2021jfp}. The standard formula for the electron structure functions, $D$, includes three terms presented by Eq.~(\ref{esf3}). The functions $D^\gamma$ in the formalism of the electron structure functions that correspond to the initial and final state radiation are identical, but respective functions obtained and used in \cite{Liu:2021jfp} (they are refereed as the universal lepton distribution and lepton fragmentation functions) are not due to the difference in the factor under the leading log in (2.18) and (2.20) of \cite{Liu:2021jfp}. We note, that this difference does not affect the leading log part of the total RC. Furthermore, the functions $D$ presented in \cite{Liu:2021jfp} are not complete because they do not include the effects of collinear  electron pair production (i.e., $D^{e^+e^-}_N=0$ and $D^{e^+e^-}_S=0$) and the effects of multiple photon emission is presented in the lowest order only, i.e., the function $D^\gamma$ contains only the term with the $\delta$-function and first term in the sum over $k$ in Eq. (8) of \cite{AAM2004}.

 Finally, several contributions unavoidable when we calculate the total RC exactly or in the leading log approximation are not presented in the formulae of ref. \cite{Liu:2021jfp}. These include the elastic and exclusive radiative tails for inclusive and semi-inclusive RC (as we partly discussed above) and the effect of vacuum polarization for both processes.

The availability of both exact and leading log formulae allowed us to perform detailed comparison of RC calculated using both approaches. We found that generally the leading log approximation gives the main contribution in the kinematics of modern JLab measurements. The factor $\log{Q^2}/{m^2}$ is of order 15 for JLab energies, so leading log approximation provides a reasonable approximation in the broad range of kinematics. However, we also detected the regions where the next-to-leading correction cannot be avoided, e.g., at the region near $\phi_h=180^o$ and small $z$. The role of the next-to-leading terms is expected to be more important in the case of polarization measurement. For asymmetries the leading log terms are (partly) factorized, so can have a tendency for cancellation in the numerator (spin-dependent part of the cross section) and denominator (unpolarized part of the cross section).

We note that the formula (\ref{lonlo}) gives an idea on how to extract the leading log results numerically using the available code for exact RC computation. We need to obtain the results for $\sigma '_{RC}$ calculated for an artificial value of the electron mass $n\;m$, (where $n$ is an arbitrary value, e.g.,mass $n\;m$, $n=10$) in addition to the results calculated using the regular value of $m$. Since both $A$ and $B$ are independent on the electron mass, the value of $A$ can be obtained as $\sigma '_{RC}-\sigma_{RC}=\bigl(\log(Q^2/n^2m^2)-\log(Q^2/m^2)\bigr)A=-\log(n^2)A$. This approach provides a tool allowing us to test both leading log codes and codes that is based on the exact formulae.

\appendix
\section{Treatment of the infrared divergence}
\label{app}
According to the Bardin-Shumeiko approach \cite{BSh} the infrared divergence in (\ref{dsrsp}) has to be extracted using an identical transformation:
\ba
d\sigma^R_{s,p}=d\sigma^R_{s,p}-d\sigma^{IR}_{s,p}+d\sigma^{IR}_{s,p}=d\sigma^{F}_{s,p}+d\sigma^{IR}_{s,p},
\label{AppA1}
\ea
where 
\ba
d\sigma^{IR}_s&=&\frac{\alpha}{2\pi^2}d\sigma_B\frac {1}{(1-z_1)kk_1}\frac {d^3k}{k_0},
\nn[-2mm]
\nn
d\sigma^{IR}_p&=&\frac{\alpha}{2\pi^2}d\sigma_B\frac {z_2}{(1-z_2)kk_2}\frac {d^3k}{k_0}.
\label{irll}
\ea
The transformation (\ref{AppA1}) is performed in the dimensional regularization. The terms $d\sigma^{F}_{s,p}$ obtained as the result of subtraction of (\ref{irll}) are infrared free, so can be further dealt with in the regular four-dimensional space. The methods describing in Section \ref{subsecCollpoles} allow to represent these terms in the form: 
\begin{widetext}
\ba
\sigma^F_s&=&\sigma^R_s-\sigma^{IR}_s
\nn
&=&\frac\alpha {2\pi }l_m \int\limits_{z_1^m}^1
dz_1\Biggl(
\frac{1+z_1^2}{1-z_1}\frac{p_{ls}S_x^{2}}{p_l(z_1S-X)^2}
\sigma_B(z_1S,z_1Q^2,x_s,z_s,p_{ts},\cos \phi_{hs})
-\frac{2}{1-z_1}\sigma_B(S,Q^2,x,z,p-_{t},\cos \phi_{h})
\Biggr),
\nonumber\\
\sigma^F_p&=&\sigma^R_p-\sigma^{IR}_p
\nn
&=&\frac\alpha {2\pi }l_m \int\limits_{z_2^m}^1 \frac{dz_2}{z_2}
\Biggl(
\frac{1+z_2^2}{z_2(1-z_2)}\frac{p_{lp}S_x^2}{p_l(S-X/z_2)^2}
\sigma_B(S,z_2^{-1} Q^2,x_p,z_p,p_{tp},\cos \phi_{hp})
-
\frac{2}{1-z_2}\sigma_B(S,Q^2,x,z,p_{t},\cos \phi_{h})
\Biggr),
\nn
\label{sfsp}
\ea
\end{widetext}
where the lowest limits of integration are defined by Eqs.~(\ref{z12m}). 

The remaining terms $d\sigma^{IR}_{s,p}$ are infrared divergent, so all further manipulations with them have to be performed in the dimensional regularization.  Using methods of Appendix C of \cite{AI2019} we obtain the resulting expressions in the leading order:
\ba
&\displaystyle
\sigma^{IR}_s+\sigma^{IR}_p
=\frac{\alpha}{\pi}\delta_{IR}\sigma_B(S,Q^2,x,z,p_{t},\cos \phi_{h})
\nn
&\displaystyle
=
\frac{\alpha}{\pi}\Biggl[l_m\biggl(2P_{IR}+2\log \frac m \nu+\log\frac{(p_x^2-M_{th}^2)^2}{S^\prime X^\prime}\biggr)
\nn
&\displaystyle
+\frac 12l_m^2\Biggr]\sigma_B(S,Q^2,x,z,p_{t},\cos \phi_{h}).
\ea

Both the infrared divergence $P_{IR}$ term and the term containing the square of $l_m$ cancel in the sum
with the corresponding vertex contribution
that can be obtained from Eq.~(50) of \cite{AI2019} in the limit $m\to 0$:
\ba
\delta_{\rm vert}=l_m\biggl (\frac 32-2P_{IR}-2\log \frac m \nu\biggr)-\frac 12l_m^2.
\ea

Summing up $\sigma^F_{s,p}$ defined by Eq.~(\ref{sfsp}), $\sigma^{IR}_{s,p}$, $\alpha/\pi\delta_{\rm vert}\sigma_B(S,Q^2,x,z,p_{t},\cos \phi_{h})$ and
vacuum polarization $\alpha/\pi\delta_{\rm vac}^l(Q^2)\sigma_B(S,Q^2,x,z,p_{t},\cos \phi_{h})$ we can find that radiative corrected cross section in leading approximation reads
\ba
\sigma^{in}_{1L}&=&\Biggl[1+
\frac{\alpha }{\pi }(\delta_{VR}
+\delta_{\rm vac}^l(Q^2)
)\biggr]\sigma_{B}(S,Q^2,x,z,p_t,\phi_h)
\nn&&        
+\sigma^F_s
+\sigma^F_p,
\label{srvll}
\ea
where
\ba
\delta_{VR}=\delta_{IR}+\delta_{\rm vert}=l_m\log\frac{(p_x^2-M_{th}^2)^2}{S^\prime X^\prime}
+\frac 32l_m,
\label{dvrll}
\\\nonumber
\ea
and $\delta_{\rm vac}^l(Q^2)$ is defined by Eq.~(\ref{dvac}).

The expression for $\sigma^{in}_{1L}$ can be explicitly presented in terms of the splitting function (\ref{spf}):
\begin{widetext}
\ba
\sigma^{in}_{1L}=
\Biggl[1+
\frac{\alpha }{\pi }\delta_{\rm vac}^l(Q^2)\biggr]\sigma_B(S,Q^2,x,z,p_t,\phi_h)
+\frac{\alpha}{2\pi}l_m
\Biggl[\int\limits_{z_1^m}^1 dz_1P(z_1)
\frac{p_{ls}S_x^2}{p_l(z_1S-X)^2}
\sigma_B(z_1S,z_1Q^2,x_s,z_s,p_{ts},\cos \phi_{hs})
\nn
+\int\limits_{z_2^m}^1 dz_2\frac{P(z_2)}{z_2^2}
\frac{p_{lp}S_x^2}{p_l(S-X/z_2)^2}
\sigma_B(S,Q^2/z_2,x_p,z_p,p_{tp},\cos \phi_{hp})
\Biggr].
\nn
\label{srvll2}
\ea
\end{widetext}
The explicit expression for $\sigma^{in}_{1L}$ is given in Eq.~(\ref{sll1}).

\bibliography{mybibfile}

\end{document}